# Sparse Identification of Nonlinear Dynamics Enhanced by Ensemble Learning, Multi-Step Prediction Evaluation, Elite Strategy, and Classification Techniques for Applications to Industrial Systems

Shuichi Yahagi*[1], Ansei Yonezawa[2], Hiroki Seto[3], Heisei Yonezawa[4], and Itsuro Kajiwara[4]

[1] *Department of Mechanical Engineering, Tokyo City University, 1-28-1 Tamazutsumi, Setagaya-ku, Tokyo, 158-8557, Japan. yahagis@tcu.ac.jp*

[2] *Department of Mechanical Engineering, Kyushu University, 744 Motooka, Nishi-ku, Fukuoka 819-0395, Japan.*

[3] *6th Research Department, ISUZU Advanced Engineering Center Ltd., 8 Tsutidana, Fujisawa-shi, Kanagawa 252-0881, Japan.*

[4] *Division of Mechanical and Aerospace Engineering, Hokkaido University, N13, W8, Kita-ku, Sapporo, Hokkaido 060-8628, Japan.*

Abstract— This paper proposes a sparse identification of nonlinear dynamics (SINDy) with control and exogenous inputs for highly accurate and reliable prediction. Although SINDy is recognized as a remarkable approach for identifying nonlinear systems, several challenges remain. Its application to industrial systems remains limited, and multi-step predictions are not guaranteed due to overfitting and noisy data. This phenomenon is often caused by the increase in basis functions resulting from the extension of coordinates, such as time-delay embedding. To address these problems, this study proposes an emphasized SINDy framework by integrating ensemble-learning, multi-step prediction evaluations, elite strategy, and classification techniques (EMEC-SINDy), while preserving convex optimization. The proposed method employs library bagging and extracts elites with an R-squared greater than 90%. Then, clustering is performed on the surviving elites because physically motivated basis functions are not always available, and the elites obtained do not always have similar basis functions. After the classification, discrete model candidates are obtained by taking the mean of each classified elite. Finally, the best model is selected. Simulation results demonstrate that EMEC-SINDy significantly outperforms original SINDy approaches in multi-step prediction accuracy under noisy conditions, validating its applicability to the diesel engine airpath system, which is known as a complex and highly coupled nonlinear multi-input multi-output system.

# 1. Introduction

System identification is essential for analysis, controller design, and the prediction of dynamical systems. In linear systems, traditional system identification methods, such as the Auto-Regressive with eXogenous input (ARX) model and Numerical Algorithms for Subspace State Space System Identification (N4SID) (Van Overschee and De Moor, 1994), are effective. However, those methods suffer from obtaining desirable models of complex industrial systems with strong nonlinearities. Over the past decades, extensive research has been conducted on the system identification of nonlinear dynamics (Kerschen et al., 2006; Noël and Kerschen, 2017; Schoukens and Ljung, 2019). Numerous studies have been carried out to develop physical models for nonlinear systems in various communities, including the field of machine learning and control engineering. Traditional machine learning methods, such as deep learning and reinforcement learning, have been presented to develop the desired controllers and models for complex systems (Jin et al., 2025; Li et al., 2024; Nabeel et al., 2024; Peng et al., 2024). However, computational and learning costs for these approaches are problematic. Recently, the data-driven science community has introduced several alternative methods distinct from deep learning based on neural networks, such as sparse identification of nonlinear dynamical systems (SINDy) (Brunton et al., 2016a), dynamic mode decomposition (DMD) (Lv et al., 2023; Peter and Joern, 2008; Schmid, 2010), and Koopman analysis (Bakhtiaridoust et al., 2023; Brunton et al., 2022; Korda and Mezić, 2018). Among these, SINDy facilitates sparsity-promoting modeling. This helps mitigate overfitting and lowers computational effort. A previous study demonstrated that SINDy outperforms neural networks in terms of computational efficiency and exceeds DMD in terms of modeling accuracy (Kaiser et al., 2018). Due to these advantages, SINDy is well-suited for model predictive control (MPC) (Fasel et al., 2021), which relies on real-time optimization using an identified system model to achieve desired control performance. MPC is particularly attractive due to its applicability to nonlinear systems, ability to handle constraints, and flexibility in cost function design. Recently, machine learning-based MPC frameworks have attracted significant attention (Ren et al., 2022; Bonassi et al., 2022; Norouzi et al., 2023). For instance, studies (Huang et al., 2024a, 2024b, 2023) have presented MPC approaches based on error-triggered sparse identification, fuzzy neural network, and long short-term memory (LSTM) to address nonlinear systems with wide operating ranges. Additionally, numerous applications to industrial systems have been reported (Schwenzer et al., 2021). However, the performance of MPC is highly dependent on the accuracy of the identified models, and low computational cost is essential for implementation on embedded controllers. Therefore, SINDy-based approaches are suitable for identifying models for MPC.

Various studies have been conducted on the applicability of SINDy to nonlinear dynamical systems. Previous literature has demonstrated the high effectiveness of SINDy for susceptible-exposed-infectious-removed epidemic models and Lorenz equations, both of which exhibit nonlinear and multivariable dynamics (Brunton et al., 2016a, 2016b; Fasel et al., 2022, 2021). Several algorithms inspired by SINDy have also been utilized in physics (Champion et al., 2020), fluid dynamics (Brunton et al., 2016a), biology (Mangan et al., 2016), COVID-19 research (Jiang et al., 2021), and chemical processes to identify governing equations and dynamical systems (Brunton et al., 2016a, 2016b; Fasel et al., 2022, 2021). Among them, there are numerous applications in chemical processes, including hydraulic fracturing (Narasingam and Sang-Il Kwon, 2018), a continuous stirred tank reactor (Bhadriraju et al., 2019), an isothermal batch reactor (Abdullah et al., 2021), and a distillation column (Subramanian et al., 2021). In this way, SINDy has achieved a certain degree of success. Furthermore, many SINDy-based algorithms have been proposed, including ensemble-SINDy (Fasel et al., 2022), dropout-SINDy (Abdullah et al., 2022), FE (feature engineering)-SINDy (França et al., 2022), SINDy-SA (sensitivity analysis) (Naozuka et al., 2022), SINDy with the Bayesian approach (Chatterjee et al., 2023; Fuentes et al., 2021; Zhu et al., 2022), kSINDYc (key term-based SINDYc) (Xavier et al., 2024), iNeural-SINDy (Neural networks and integrating schemes assisted SINDy) (Forootani et al., 2025), and SINDy with Akaike information criterion (Dong et al., 2023). However, most existing approaches focus on systems in which the physical dynamics can be explicitly formulated and represented as linear combinations of candidate basis functions in a predefined library. Previous research has not sufficiently explored the application of SINDy to systems whose characteristics are not explicitly known, such as complex industrial systems. This suggests that, although many excellent methods have been proposed, their application to industrial systems remains limited. Ensuring desirable predictive performance for complex real-world systems under noisy conditions presents several challenges. These include the appropriate selection of candidate basis functions that constitute the library, and the risk of overfitting due to library expansion and the presence of noise (Abdullah et al., 2022; Fasel et al., 2022; França et al., 2022). Unlike simple systems, where mathematical expressions can be derived from physical laws, constructing the library based on physical knowledge is difficult for industrial systems. While extension of coordinate (i.e., phase space), such as time-delay embedding, is effective for expressing complex systems, the library increases (Hajiloo et al., 2018; Wang et al., 2023). Previous research (Kiser et al., 2023) has pointed out that a large number of library sets can lead to multicollinearity and overfitting. Thus, it isn't easy to realize accurate predictions for industrial systems. From the above, sparse identification for industrial systems remains an open problem. In this study, we aim to develop a sparse identification methodology tailored to industrial systems.

The system considered in this paper is a diesel engine airpath system, which exhibits nonlinear and multivariable dynamics. Although various approaches have been proposed (Aran and Unel, 2020; Hirata et al., 2019; Ishizuka et al., 2017; Moriyasu et al., 2019; Xie et al., 2016), the control and modeling of diesel engine airpath systems remain challenging. In the model-based approach, the study (Hirata et al., 2019) presents the feedforward controller designed using a nominal model derived from physical modeling. In (Moriyasu et al., 2019), neural network modeling and MPC are presented. However, physical modeling suffers from limited accuracy and difficulties in parameter identification. In neural network modeling, computational and learning costs are problematic. In contrast, the approach that does not rely on system models is presented due to the difficulty of modeling diesel engines. The study (Ishizuka et al., 2017) introduces the model-free adaptive PID control based on simultaneous perturbation stochastic approximation (SPSA). Literature (Aran and Unel, 2020; Xie et al., 2016) proposes disturbance observer-based control. However, ensuring closed-loop stability is difficult without an accurate plant model. In addition, the performance of the designed controller cannot be evaluated without conducting real-world experiments, and prior desk-based validation is not feasible. To address these problems, SINDy-based approaches are considered for the diesel engine airpath system.

This paper introduces a robust SINDy with control and exogenous inputs to obtain ordinary difference equations to realize multi-step prediction. Multi-step prediction is essential from the perspective of simulation plant utilization and MPC applications. For complex, highly coupled, nonlinear multi-input, multi-output (MIMO) systems such as diesel engine airpath systems (Hu et al., 2018; Moriyasu et al., 2019), it is effective to increase the expressive capability of SINDy by introducing time-delay coordinates into the library. However, in this setting, the increase in the number of basis functions (libraries) leads to overfitting, making it difficult to obtain accurate models. To address this issue, we propose an emphasized SINDy framework by integrating ensemble-learning, multi-step (long-term) prediction evaluations, elite strategy, and classification techniques (EMEC-SINDy). First, multiple libraries are selected randomly, and each coefficient matrix of SINDy is identified by convex calculation, resulting in ordinary difference equations for multiple libraries. Next, by performing the long-term (multi-step) prediction for each model with given initial states and given inputs, the coefficient of determination (R-squared, $R^2$) is obtained. Using the results of the R-squared, we determine the surviving elites. That is, models that fail in long-term prediction are discarded, and models that achieve long-term prediction survive as the elites. Here, the elite models obtained do not always have similar basis functions. Thus, after classification is applied to the elites, the model is finally obtained by taking the mean values of the classified elites. The remarkable advantage of the proposed approach is that the desired ordinary difference equation for realizing multi-step prediction can be obtained by solving a convex problem. This convex formulation helps reduce computational complexity. Additionally, we apply the proposed method to the diesel engine airpath system with nonlinear MIMO characteristics. As far as the authors know, the application of SINDy to airpath systems of diesel engines has not been studied except for the study (Yahagi et al., 2025). We summarize the contributions of this paper as follows:

- This paper presents EMEC-SINDy with control and exogenous inputs for modeling industrial systems under noisy conditions. To obtain highly accurate predictive models for complex systems such as industrial systems, the number of candidate basis functions included in the library is typically increased to enhance the model's ability to capture complex dynamics. However, in conventional methods, the expansion of the candidate basis functions and the presence of noise can lead to overfitting. The proposed method addresses this issue by integrating elite gathering, ensemble learning based on multi-step prediction evaluation, and clustering techniques. While conventional ensemble learning alone does not sufficiently promote sparsification, the proposed method incorporates clustering to enhance sparsity and improve model robustness. Furthermore, this paper treats SINDy with control and exogenous inputs, whereas many existing studies focus primarily on autonomous systems without inputs.
- Most previous SINDy-based research has often focused on systems where differential equations are clearly or explicitly expressed, and its application to more complex nonlinear systems, such as industrial systems, has been limited. In this paper, we implement the proposed EMEC-SINDy in a diesel engine airpath system using a library that introduces extended coordinates to enhance the expressiveness of nonlinearities. As this system is a challenging nonlinear MIMO system and existing research aside from reference (Yahagi et al., 2025) is limited, this study offers novel insights for both industrial engineers and researchers.

This paper is structured as follows: Section 2 provides an overview of the airpath system in internal combustion engines, which is the focus of our modeling efforts. Since data-driven modeling relies on data rather than a physical model, we will not delve into detailed physical formulas but will instead present a general overview of the system. Section 3 outlines the proposed data-driven modeling approach, which aims to ensure accurate multi-step predictions for nonlinear systems even in the presence of noise. In Section 4, the proposed method, EMEC-SINDy, is evaluated through its application to both a simple illustrative example and a diesel engine airpath system under noisy conditions, using numerical simulations. The basic characteristics of the proposed method are first investigated through illustrative examples. Subsequently, its performance is assessed in the diesel engine airpath system. Finally, Section 5 offers a summary of this paper.

# 2. Target system

## 2.1. System overview

The 4-cylinder diesel engine's airpath system, depicted in Fig. 1. This system has a variable geometry turbocharger (VGT) and an exhaust gas recirculation (EGR) system. The VGT adjusts the intake manifold pressure by varying the vane spacing. Narrowing the vane spacing increases flow velocity, providing a supercharging effect at low speeds. At high speeds, the vane opening angle is adjusted to improve exhaust flow. The EGR system regulates the oxygen content in the cylinder by blending fresh air with exhaust gas. This helps suppress the formation of harmful nitrogen oxides (NOx), which are typically produced under high-temperature conditions.

The gas flow in the system proceeds as follows. Ambient air is compressed and cooled by a compressor and an intercooler, respectively, before passing through the intake throttle and entering the intake manifold. From there, the gas enters the cylinder, where combustion occurs. The combustion gases are then directed into the exhaust manifold. At this point, the exhaust flow is split into two routes, with one portion being recirculated by the EGR system. The EGR valve is adjusted to regulate the oxygen content inside the intake manifold. This setup is referred to as a high-pressure EGR system, in which relatively hot exhaust gas from the upstream side of the turbine is returned to the intake manifold. The other portion is the exhaust gas through the turbine. The gas after combustion passes through the exhaust manifold and works the turbine. The turbine vane angle is manipulated to control the turbine rotation speed. In this study, the derivation of physical modeling will be omitted since this paper adopts a data-driven approach. For details, refer to (Kaneko et al., 2019; Wahlström and Eriksson, 2011; Yahagi et al., 2025).

## 2.2. System dynamics

We describe the input-output signals of the system. Table I summarizes the symbols and their corresponding parameter names. The system outputs, denoted by $y_1 \in \mathbb{R}$ and $y_2 \in \mathbb{R}$, represent the intake manifold pressure [kPa] and the EGR rate [%], respectively. Here, the intake manifold pressure is also referred to as the boost pressure. The control inputs, $u_1 \in \mathbb{R}$ and $u_2 \in \mathbb{R}$, correspond to the VGT vane position [%, closed] and the EGR valve position [%, open], respectively. The exogenous inputs, denoted by $d_1 \in \mathbb{R}$ and $d_2 \in \mathbb{R}$, correspond to the fuel injection quantity [mm³/st] and engine revolution [rpm], respectively. These values are determined by the vehicle conditions resulting from the driver's actions. The symbols $u \in \mathbb{R}^2$, $d \in \mathbb{R}^2$, and $y \in \mathbb{R}^2$ are the control input vector, exogenous input vector, and output vector, respectively. Therefore, a dynamics system is expressed as

$$y^{(n_y)}(t) = f_p\left(\phi_y(t), \phi_u(t), \phi_d(t)\right) \tag{1}$$

with

$$\begin{aligned} \phi_y(t) &= [y^{(n_y-1)}(t), y^{(n_y-2)}(t), \dots, y(t)] \\ \phi_u(t) &= [u^{(n_u-1)}(t), u^{(n_u-2)}(t), \dots, u(t)] \\ \phi_d(t) &= [d^{(n_d-1)}(t), d^{(n_d-2)}(t), \dots, d(t)] \end{aligned} \tag{2}$$

where $f_p$ is the nonlinear function of the system; $n_u \in \mathbb{Z}$, $n_d \in \mathbb{Z}$, and $n_y \in \mathbb{Z}$ are the order of control input, exogenous input, and output of the system, respectively; $t$ is the time. It is noted that many unmeasurable states are included in the system, although this section focuses on the input-output relationship. Note that the airpath system described above exhibits several challenging characteristics (Kaneko et al., 2019; Wahlström and Eriksson, 2011; Yahagi et al., 2025). These include constraints on control inputs (i.e., actuators), significant interference in the air flow caused by the operation of the VGT and EGR, and operating-point-dependent characteristics caused by the fuel injection amounts and engine revolutions. In this paper, EMEC-SINDy is proposed to enable modeling and control of this challenging system. It provides a discrete ODE model capable of multi-step prediction under noisy conditions.

Table I. Symbols and their parameter names.

| Symbol | | Parameter name | Unit |
|---|---|---|---|
| $y$ | $y_1$ | Boost pressure (intake manifold pressure) | Pa |
| | $y_2$ | EGR ratio | % |
| $u$ | $u_1$ | VGT vane closure | % |
| | $u_2$ | EGR valve opening | % |
| $d$ | $d_1$ | Fuel injection amount | mm³/st |
| | $d_2$ | Engine revolution | rpm |

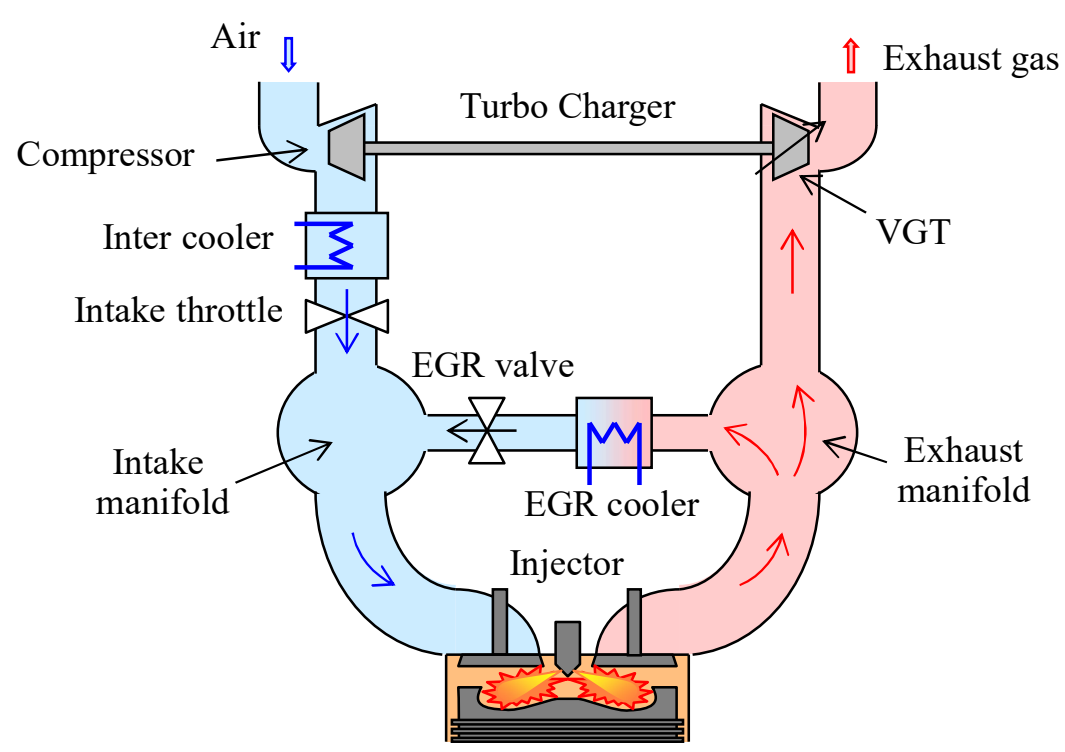

Fig. 1. Diesel engine airpath system diagram.

# 3. Proposed method

## 3.1.Original SINDy with control and exogenous inputs

In the data science field, SINDy (Brunton et al., 2016b; Fasel et al., 2021) has been introduced for understanding nonlinear dynamics. SINDy offers several appealing features, such as superior computational cost due to sparsity, effective learning capabilities due to convexity, accurate modeling, and suitability for complex systems. We explain the original SINDy framework, incorporating control and exogenous inputs, to model the system described in Section 2. The formulation is introduced in a discrete-time form since the controller design, including MPCs, is generally conducted using a discrete model (see Appendix). If the model is described in continuous time, integration, such as the Runge-Kutta method is necessary in MPC implementation. In addition, the signals are obtained by the zero-order holder (ZOH), i.e., the digital signals are available. Thus, this study considers the discrete nonlinear dynamical system:

$$x(t+1) = f\big(x(t), u(t), d(t)\big) \tag{3}$$

where $f\colon \mathbb{R}^n \times \mathbb{R}^l \to \mathbb{R}^n$ is the nonlinear function of the system; $x(t) = [x_1(t) \quad x_2(t) \quad \cdots \quad x_n(t)\ ]^T \in \mathbb{R}^n$, $u(t) = [u_1(t) \quad u_2(t) \quad \cdots \quad u_l(t)\ ]^T \in \mathbb{R}^l$, and $d(t) = [d_1(t) \quad d_2(t) \quad \cdots \quad d_q(t)]^T \in \mathbb{R}^q$ are the $n$ dimensional state vector, $l$ dimensional control input vector, and $q$ dimensional exogenous input vector, respectively. It is noted that the state vector $x$ and exogenous input vector $d$ are measurable. Then, the snapshot data vectors of each variable for the length of data, $m$, are given as

$$X = \begin{bmatrix} | & | & | & | \\ x(1) & x(2) & \cdots & x(m) \\ | & | & | & | \end{bmatrix} \in \mathbb{R}^{n\times m} \tag{4}$$

$$X^+ = \begin{bmatrix} | & | & | & | \\ x(2) & x(3) & \cdots & x(m+1) \\ | & | & | & | \end{bmatrix} \in \mathbb{R}^{n\times m} \tag{5}$$

$$\Gamma = \begin{bmatrix} | & | & | & | \\ u(1) & u(2) & \cdots & u(m) \\ | & | & | & | \end{bmatrix} \in \mathbb{R}^{l\times m} \tag{6}$$

$$D = \begin{bmatrix} | & | & | & | \\ d(1) & d(2) & \cdots & d(m) \\ | & | & | & | \end{bmatrix} \in \mathbb{R}^{q\times m}. \tag{7}$$

In the formulation of SINDy, the system dynamics (3) are represented as a linear combination of the basis function (i.e., library) $\vartheta\colon \mathbb{R}^n \times \mathbb{R}^l \times \mathbb{R}^q \to \mathbb{R}^p$, associated with a coefficient matrix $\Xi \in \mathbb{R}^{n\times p}$, as follows:

$$x(t+1) = \Xi\vartheta^T(x(t), u(t), d(t)) \tag{8}$$

with

$$\Xi = \begin{bmatrix} - & \xi_1 & - \\ - & \xi_2 & - \\ & \vdots & \\ - & \xi_n & - \end{bmatrix} \tag{9}$$

$$\vartheta\big(x(t), u(t), d(t)\big) = \begin{bmatrix} \vartheta_1\big(x(t), u(t), d(t)\big) \\ \vartheta_2\big(x(t), u(t), d(t)\big) \\ \vdots \\ \vartheta_p\big(x(t), u(t), d(t)\big) \end{bmatrix}^T. \tag{10}$$

As an example, the basis function $\vartheta$ is defined as

$$\begin{aligned} \vartheta(x, u, d) = [1^T \quad x^T \quad u^T \quad d^T \quad (x \otimes x)^T \quad (x \otimes u)^T \quad (x \otimes d)^T \quad \cdots \\ \sin(x)^T \quad \sin(u)^T \quad \sin(x \otimes x)^T \quad \sin(x \otimes u)^T \quad \sin(x \otimes d)^T \quad \cdots] \end{aligned} \tag{11}$$

where the symbol $\otimes$ denotes the Kronecker product, and the time index $t$ is omitted for simplicity. The sparse coefficient matrix $\Xi$ is then identified through SINDy algorithm. Based on the collected snapshot data and the system dynamics formulated in the SINDy framework (8), the following equation is obtained:

$$X^+ = \Xi\Theta^T(X, \Gamma, D) \tag{12}$$

where $\Theta: \mathbb{R}^{n\times m} \times \mathbb{R}^{l\times m} \times \mathbb{R}^{q\times m} \to \mathbb{R}^{m\times p}$ is the matrix form of the library (10), defined as

$$\Theta(X, \Gamma, D) = \begin{bmatrix} -\vartheta\big(x(1), u(1), d(1)\big) - \\ -\vartheta\big(x(2), u(2), d(2)\big) - \\ \vdots \\ -\vartheta\big(x(m), u(m), d(m)\big) - \end{bmatrix}. \tag{13}$$

By incorporating a regularization term to mitigate overfitting and compress the data, the following optimization problem is formulated:

$$\xi_i = \arg\min_{\xi_i'} \|X_i^+ - \xi_i'\Theta^T(X, \Gamma, D)\|_2^2 + \mathcal{R}(\xi_i') \tag{14}$$

where $X_i^+$ represents the $i$-row component of $X^+$, and $\mathcal{R}(\xi_i)$ denotes the regularizer to promote sparsity in $\xi_i$. In this paper, $\mathcal{R}(\xi_i)$ is defined as $\lambda_0\|\xi_i\|_0$, where $\lambda_0$ is the sparsity-promoting hyperparameter. This hyperparameter is empirically tuned to achieve the best estimation of dynamics. In this optimization problem, a sparsified coefficient vector $\xi_i$ is obtained using STLS (sequentially thresholded least-squares) (Brunton et al., 2016a; Kaiser et al., 2018) or LASSO (least absolute shrinkage and selection operator) regression (Tibshirani, 1996).

*Remark 1.* The original SINDy is expected to be applied to nonlinear MIMO systems; however, multi-step prediction is not guaranteed. In the simulation section, we show that the original SINDy may provide a model in which the one-step prediction is possible, but the multi-step prediction is not feasible. Thus, we present the new algorithm of SINDy for realizing multi-step predictions in the next section.

### 3.2. Time-delay coordinate

The input-output dynamics is introduced for the targeted system in Section 2. Herein, note that many states are included in $x(t)$. Although it is assumed that all states are measurable in the original SINDy, many of them are not measurable in real systems. Indeed, the measurable states are only outputs in the airpath system. Thus, we need to extend the coordinates (phase space) from measurable states given as

$$x_m(t) = [x^T(t) \quad x^T(t-1) \quad \cdots \quad x^T(t-\sigma_x)]^T \in \mathbb{R}^{n\sigma_x} \tag{15}$$

where $\sigma_x$ denotes the user-defined delay order of the states. Then, a dynamic system with embedded delay time is expressed as

$$\begin{cases} x_m(t+1) = f\big(x_m(t), u(t), d(t)\big) \\ \quad\quad y(t) = h\big(x_m(t)\big) = x(t) \end{cases} \tag{16}$$

A dynamic model can be obtained by replacing $x$ with $x_m$, and solving the optimization problem derived in the previous section. Time-delay coordinate has been proposed in various studies (Hajiloo et al., 2018; Wang et al., 2023).

### 3.3. Model validation

In many regression problems, including Nonlinear AutoRegressive model with eXogenous inputs (NARX) and original SINDy, the regression coefficients are obtained by evaluating one-step-ahead predictions. In nonlinear systems, multi-step prediction may not be achievable even when accurate one-step prediction is achieved, unlike linear systems (Xiao et al., 2023), because small errors can compound over time, resulting in poor long-horizon predictions (Somalwar et al., 2025). It is well known that multi-step-ahead prediction and dynamic modeling are significantly more challenging than one-step-ahead prediction (Menezes and Barreto, 2008). Therefore, we evaluate both one-step and multi-step predictions. The one-step prediction is defined as

$$\begin{cases} \hat{x}(t+1) = f\big(x(t), u(t), d(t)\big) \\ \quad\quad \hat{y}(t) = h(\hat{x}(t)) \end{cases} \tag{17}$$

where $\hat{y}$ is the predicted value of $y$. The one-step ahead is predicted from the given input and output data. The multi-step prediction is defined as

$$\begin{cases} \hat{x}(t+1) = f\big(\hat{x}(t), u(t), d(t)\big) \\ \quad\quad \hat{y}(t) = h(\hat{x}(t)) \end{cases} \tag{18}$$

with $\hat{x}(0) = x(0)$. The multi-step ahead is predicted from the given input and the predicted past output. It is also known that the evaluation of multi-step prediction is also important in terms of MPC.

In model validation, the modeling accuracy is evaluated using the coefficient of determination, $R^2$, of $y$ defined as

$$R^2 = 1 - \frac{\sum_{t=1}^{m}\left(y(t) - \hat{y}(t)\right)^2}{\sum_{t=1}^{m}(y(t) - \bar{y})^2} \tag{19}$$

where $\bar{y}$ is the mean value of the $m$ data (Haber and Unbehauen, 1990; Liu et al., 2017). $R^2$ equal to 1.0 indicates that the identified model best fits the target system. From an engineering perspective, an R-squared score is acceptable when it reaches a value from 0.9 to 1.0 (Schaible et al., 1997). A negative R-squared value means that the inferred model has a very low ability to represent an equivalent dynamical system. To evaluate both sparsity and prediction performance, we also introduce Akaike Information Criterion (AIC) (Bishop, 2006; Dong et al., 2023; Rubio-Herrero et al., 2022) and Bayesian Information Criterion (BIC) (Bishop, 2006):

$$AIC = -2\log L + 2N \tag{20}$$

$$BIC = -2\log L + N\log m \tag{21}$$

with

$$\log L = -\frac{nm}{2}\log\left(\frac{\sum_{t=1}^{m}\left(y(t) - \hat{y}(t)\right)^2}{nm}\right) \tag{22}$$

where $n$ is the size of output, $m$ is the number of sampling data, $N$ is the number of the model parameters.

## 3.4.The proposed algorithm

The improved ensemble-learning-based SINDy is introduced to realize multi-step ahead prediction. Fig. 2 shows the overview of the proposed method. The procedure of the proposed method is summarized in Algorithm 1. Based on the figure and the algorithm, we describe the proposed method in several parts:

(i) *Data acquisition and preprocessing*: Learning data is measured by an experiment. In data preprocessing, centering is optionally performed. Using this data and the library set by a user, we aim to obtain the SINDy model.

(ii) *Library bagging (bootstrap aggregating)* (Fasel et al., 2022): For the rich or excessive library which may cause multilinearity, library bagging is performed to promote sparsity. The number of bagging items is randomly determined, and the features are selected probabilistically. For each bagging library, the coefficient matrix is obtained by solving the optimization problem of SINDy. The optimization method adopts the STLS (Brunton et al., 2022) in this paper. The library is scaled when performing the STLS.

(iii) *Elite extraction based on multi-step prediction evaluation*: The R-squared score of multi-step predictions is evaluated for each model. This is because the STLS only evaluates one-step predictions, and multi-step predictions are not considered. For nonlinear systems, the model that realizes one-step predictions may not predict multi-step ahead. The simulation section shows this phenomenon. Thus, we should evaluate long-term (multi-step) predictions, which means ODE simulation with given inputs and initial states. Coefficient matrices of SINDy models with R-squared of 90% or higher are extracted as the elites. Note that an R-squared score from 0.9 to 1.0 is acceptable in terms of engineering application (Schaible et al., 1997).

(iv) *Classification and aggregation*: Classification is performed for the surviving elites. A clustering method, such as *k*-means and hierarchical clustering, is adopted for the coefficient matrices of the elites. Then, the average coefficient matrix for each class is calculated. While aggregating all elite models tends to hinder sparsification, aggregating the coefficient matrices within each class after clustering can enhance sparsification. Finally, the best model is selected among the models obtained for each class, based on several criteria such as R-squared, AIC (Bishop, 2006; Dong et al., 2023; Rubio-Herrero et al., 2022), and BIC (Bishop, 2006) of multi-step prediction.

*Remark 2.* In the aggregation process of ensemble-based SINDy presented in the previous study (Fasel et al., 2022), the mean value is just taken after STLS. In this previous study, inherent physical knowledge of the nonlinear system at hand seems to be taken into account (e.g., the Lotka–Volterra model). On the other hand, this paper treats a nonlinear MIMO industrial system in which it is not easy to know the features from physical knowledge. Thus, this paper introduces multi-step prediction evaluations and the classification process.

*Remark 3.* In original SINDy, $\lambda$ used in STLS is a design parameter related to sparsity. The proposed algorithm semi-automatically tunes $\lambda$. The algorithm automatically determines the value when the designer cannot find candidates. The user can also determine whether the value of its candidates is set. The guideline of hyperparameter setting is as follows: The initial value of $\lambda$ is set at the point where a significant change in sparsity is observed when the original SINDy is applied. The parameter $\Delta$ is set to a value on the order of one tenth of $\lambda$. In the simulation section, the hyperparameters are determined according to this guideline.

*Remark 4.* In Algorithm 1, a while loop is used; however, parallel computation is possible because of the bootstrap aggregating.

The proposed methodology is discussed in comparison with previous studies. Many existing approaches focus on systems in which physical dynamics can be explicitly formulated and represented as linear combinations of candidate basis functions in a predefined library, as demonstrated in the simple example (Huang et al., 2024a) presented in Section 4.1. However, the applicability of SINDy-based methods to highly nonlinear MIMO systems, such as industrial processes, has not been sufficiently explored. This means that although many excellent methods have been proposed, their application to industrial systems is still limited. The industrial system investigated in this paper presents challenges, particularly in achieving accurate multi-step prediction using original SINDy approaches. To address these issues, this paper introduces an integrated framework that combines elites' strategy, ensemble learning, and clustering. This new synergy enables robust multi-step prediction and promotes sparsity—capabilities that have not been considered in previous studies. Subsequently, we compare the proposed EMEC-SINDy with representative prior approaches, including ensemble-SINDy (Fasel et al., 2022), dropout-SINDy (Abdullah et al., 2022), SINDy-SA (Naozuka et al., 2022), and Bayesian SINDy (Chatterjee et al., 2023). Ensemble-SINDy and dropout-SINDy have not been applied to nonlinear MIMO industrial systems which are not expressed as the linear combination of a prior known basis functions, and they do not incorporate clustering to enhance sparsity. SINDy-SA employs sensitivity analysis to select the best model from multiple candidates, but it requires multiple model runs and does not utilize ensemble learning to improve robustness. In practical industrial settings, collecting multiple datasets is often labor-intensive and impractical. Bayesian SINDy is effective in handling uncertainty, but it faces challenges in convex analysis and differs fundamentally from the framework proposed in this study. In order to conduct a more detailed comparison, the next section evaluates the proposed method, EMEC-SINDy, in relation to the original SINDy and ensemble-SINDy approaches

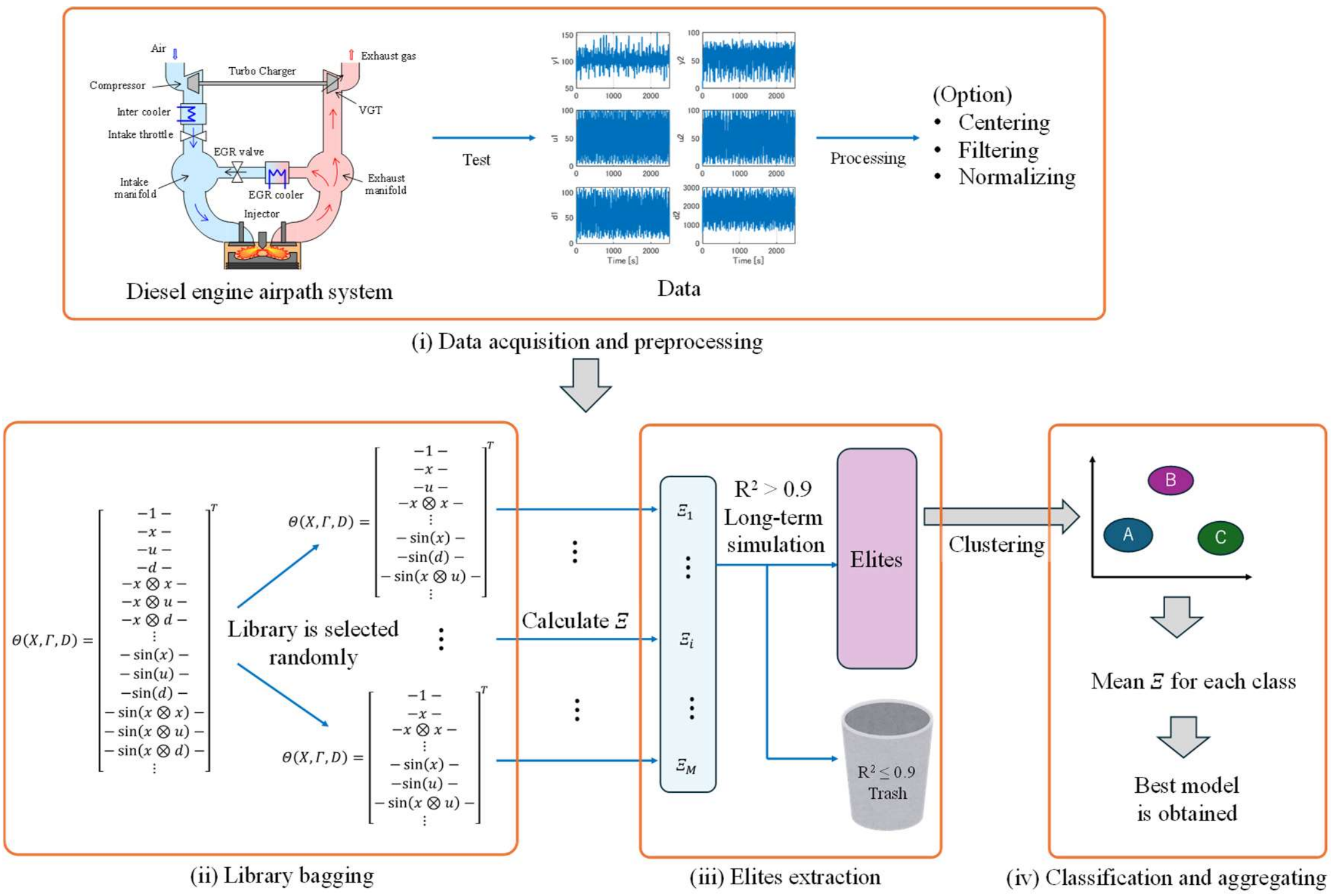

Fig. 2. Overview of EMEC-SINDy (the proposed method).

**Algorithm 1:** EMEC-SINDy (proposed method)

```
Inputs:     Library Θ
            Dataset X⁺, X, Γ, D
Outputs:    Coefficient matrix of SINDy model Ξ
Set initial λ, Δ;
While
    randomly the number and position of bagging are taken;
    compute Ξ_i using STLS with set λ;
    calculate R-squared by performing a long-term simulation with initial states and inputs;

    if R-squared ≥ 0.9
        the Ξ_i becomes elites;
    end

    if no elite is found && while statement is repeated many times
        λ = λ − Δ;
    end

    if enough elites gather
        break
    end
end
classify the elites' coefficient matrices;
mean coefficient matrices for each class;
Select the best-fit model;
```

# 4. Simulation study

In this section, a simple illustrative example and the diesel engine airpath system introduced in Section 2 are considered. The former is used to evaluate the fundamental characteristics of the proposed method and to demonstrate its validity. The latter serves as an application to an industrial system, which constitutes the primary motivation of this study. The simulation is carried out on a personal computer (PC) equipped with an Intel® Xeon® w7-3465X 2.5 GHz CPU and 128 GB of RAM. The implementation is performed using MATLAB®/Simulink. It is known that the noisy signal data may lead to reduced modeling accuracy. The noisy cases are considered in this simulation to conduct the noise robustness analysis. As well as the definition of (França et al., 2022), we added the noise to the outputs given as

$$X_v = X + \eta(Z \bigodot G) \tag{23}$$

where $\bigodot$ represents the Hadamard product; $X_v \in \mathbb{R}^{n\times m}$ is the noisy data; $\eta \in \mathbb{R}_+$ is a given noise percentage; $G \in \mathbb{R}^{n\times m}$ is Gaussian random noise with zero mean and unity variance; $Z \in \mathbb{R}^{n\times m}$ is the standard deviation matrix in which each column is the same as the standard deviation of each state.

## 4.1. Application to simple example

### 4.1.1. Simulation setting

This section presents the application of the proposed EMEC-SINDy to a simple illustrative example in order to examine its fundamental characteristics of Algorithm 1.

The following discrete-time nonlinear state-space system (Huang et al., 2024a) is considered:

$$\begin{cases} x_1(t+1) = 0.8x_1(t) - 0.2x_2^2(t) - 0.8x_1^2x_2(t) + 0.4u(t) \\ x_2(t+1) = 0.3x_1(t)x_2(t) + 0.6u(t) \\ y(t) = [x_1(t) \quad x_2(t)]^T \end{cases} \tag{24}$$

The library before the bagging is set to a cubic polynomial, defined as

$$\vartheta = [\vartheta_1 \quad \vartheta_2 \quad \cdots \quad \vartheta_i \quad \cdots \quad \vartheta_{20}] \tag{25}$$

where $\vartheta_1 = 1, \vartheta_2 = x_1, \vartheta_3 = x_2, \vartheta_4 = u, \vartheta_5 = x_1^2, \vartheta_6 = x_1x_2, \vartheta_7 = x_1u, \vartheta_8 = x_2^2, \vartheta_9 = x_2u, \vartheta_{10} = u^2, \vartheta_{11} = x_1^3, \vartheta_{12} = x_1^2x_2, \vartheta_{13} = x_1^2u, \vartheta_{14} = x_1x_2^2, \vartheta_{15} = x_1x_2u, \vartheta_{16} = x_1u^2, \vartheta_{17} = x_2^3, \vartheta_{18} = x_2^2u, \vartheta_{19} = x_2u^2, \vartheta_{20} = u^3$ . The basic and proposed SINDy employ $\lambda = 0.5$, which is the sparse promoting parameter. The number of elites is set to 1000.

#### 4.1.2. Results and discussion

Fig. 3 shows the overview of the time-series data obtained by exciting random input with an amplitude of 2.5 under a noiseless condition. The data from 0 to 100 steps are used for training, and the data from 100 to 300 steps are used as test data. As preprocessing, centering is performed. The proposed and traditional methods generate system models from this data. In the proposed method, the classification is performed for the surviving elites. *k*-means clustering is performed using the function “kmeans” in MATLAB. Fig. 4 (a) shows the results of the k-means clustering. *k*-means clustering is performed using the cityblock (Manhattan) distance metric to account for feature-wise absolute differences. To mitigate the sensitivity of *k*-means to initial centroid selection, the algorithm is executed with 1,000 random initializations, and the solution yielding the lowest total within-cluster sum of distances is selected. The value *k* is selected as the largest integer for which the silhouette score remained non-negative. Similar clusters are also obtained when applying hierarchical clustering with Ward linkage and Euclidean distance, as shown in Fig. 4 (b). Fig. 5 shows the classification results. The vertical axis represents the cluster, and the horizontal axis represents the silhouette value. We can see that proper clustering is realized because all silhouette values are positive. Various performance metrics are considered to verify the effectiveness of the proposed method. Table II indicates the R-squared of one-step and multi-step predictions for training data, the R-squared of multi-step predictions for test data, AIC and BIC of multi-step predictions for test data, and the number of elements of the identified coefficient matrix in the proposed method. Table III shows those metrics of the comparison methods, such as original SINDy and dropout-SINDy. From these performance metrics of R-squared, AIC, and BIC, it can be seen that the proposed method (Class 2) provides the best model. In addition, the identified coefficient matrix and inclusion probability (Fasel et al., 2022) are shown in Fig. 5 and Fig. 6, respectively. From these results, the proposed method provides the best model among the conventional methods (original SINDy and dropout-SINDy), in the viewpoints of R-squared, AIC, and BIC. Additionally, the number of elements *N* of the coefficients identified by dropout-SINDy (the conventional method) increases, while the proposed method achieves sparsification through clustering. Furthermore, the proposed method (Class 2) is able to extract the features of the target system more accurately than dropout-SINDy. The proposed method without clustering (denoted as "All elites" in the table) and dropout-SINDy also achieved the same results. This means that both the prediction results using one-step prediction and multi-step prediction were R-squared over 0.9, which is a result of targeting such simple nonlinear systems, or systems that can be expressed as a linear combination of the bases included in the library. In more nonlinear MIMO industrial systems, such as the intake and exhaust systems of diesel engines, which will be examined in the next section, it is difficult to know the exact basis functions, so the results of one-step prediction and multi-step prediction are significantly different. Therefore, when extracting elites, it is important to evaluate the R-squared for multi-step prediction. In the case of ensemble learning without clustering, a larger number of basis functions had low inclusion probability. On the other hand, clustering promoted sparsification and significantly increased the inclusion probability, as shown in Fig. 6. Note that the inclusion probability shows extremely small values, and the coefficient values that correspond to it could be removed; however, the values are kept since the threshold value setting is necessary. Based on the above, the proposed method is considered to have achieved the best performance across all evaluation metrics.

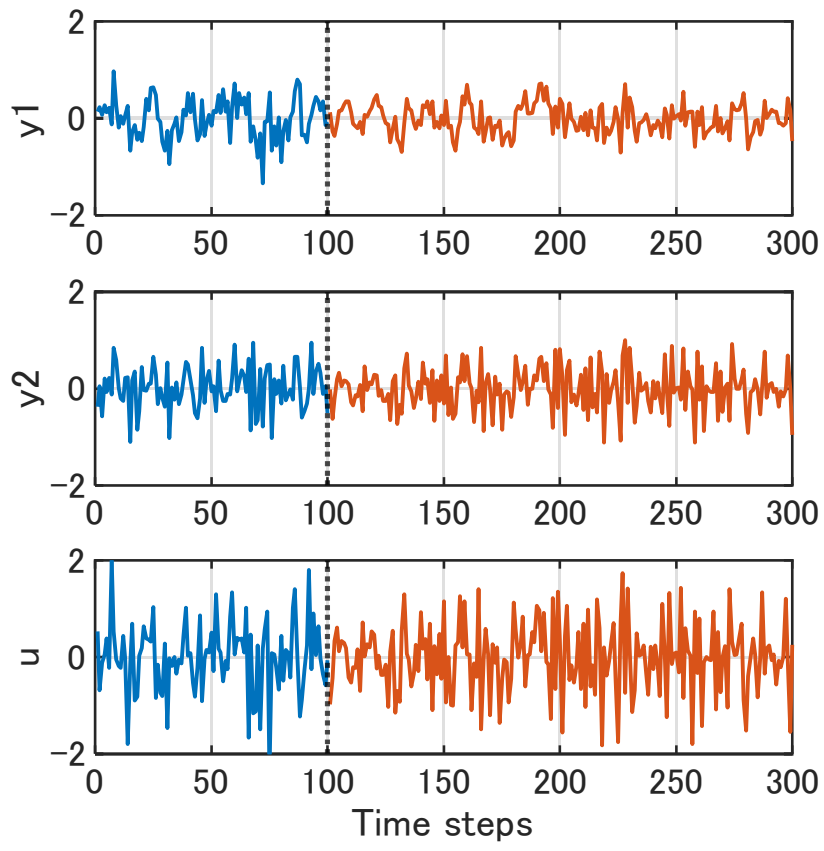

Fig. 3. Training and testing data: Time-series data from steps 0 to 100 is used for training, while data from steps 100 to 300 is used for testing.

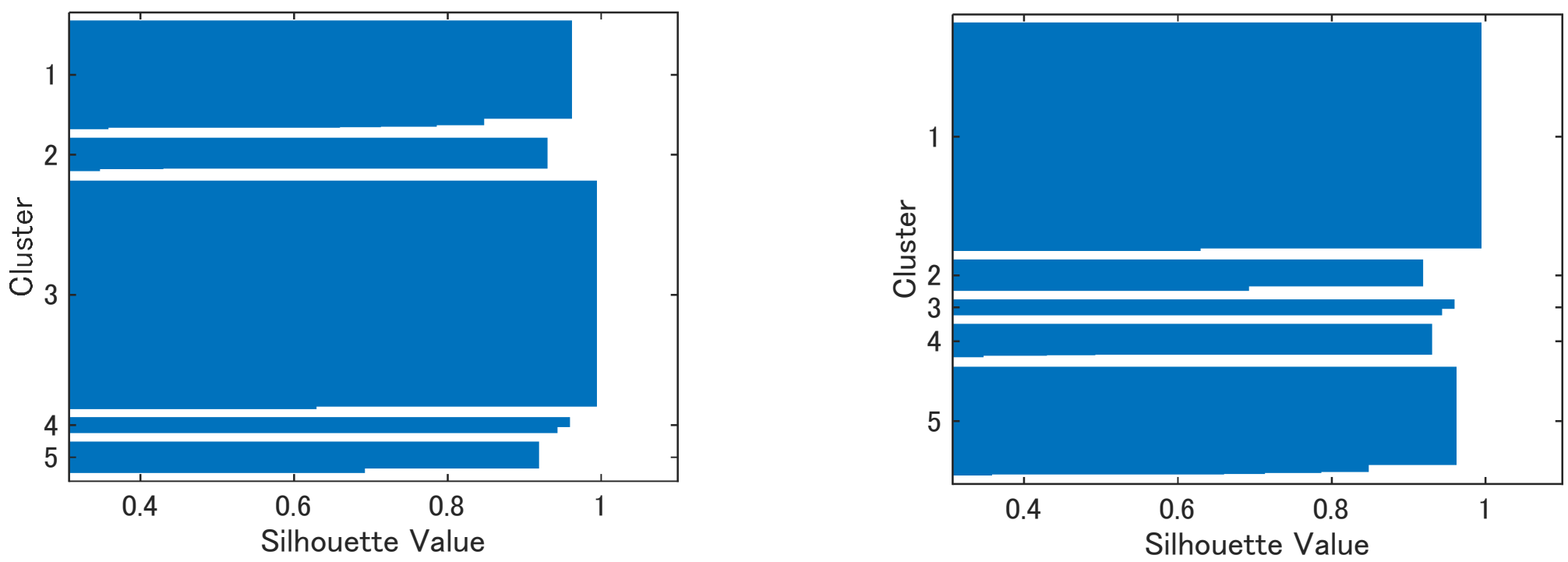

(a) *k*-means (b) Hierarchical clustering

Fig. 4. The results of *k*-means and hierarchical clustering for simple example under noiseless cases.

TABLE II. RESULTS OF PERFORMANCE METRICS IN THE PROPOSED METHOD UNDER NOISELESS CASE.

| The proposed method | One-step prediction (Training data) | | Long-term prediction (Training data) | | Long-term prediction (Test data) | | AIC | BIC | $N$ |
|---|---|---|---|---|---|---|---|---|---|
| | $y_1$ | $y_2$ | $y_1$ | $y_2$ | $y_1$ | $y_2$ | | | |
| Class 1 | 0.994 | 0.993 | 0.986 | 0.993 | 0.982 | 0.993 | −2565 | −2529 | 11 |
| Class 2 | 0.998 | 0.993 | 0.991 | 0.993 | 0.986 | 0.994 | −2648 | −2618 | 9 |
| Class 3 | 0.994 | 0.995 | 0.986 | 0.994 | 0.982 | 0.995 | −2608 | −2576 | 10 |
| Class 4 | 0.994 | 0.991 | 0.988 | 0.990 | 0.973 | 0.992 | −2464 | −2444 | 6 |
| Class 5 | 0.994 | 0.981 | 0.986 | 0.980 | 0.981 | 0.988 | −2445 | −2416 | 9 |
| All elites | 0.995 | 0.994 | 0.987 | 0.994 | 0.984 | 0.994 | −2620 | −2570 | 15 |

TABLE III. RESULTS OF PERFORMANCE METRICS IN THE TRADITIONAL METHOD UNDER NOISELESS CASE.

| The comparison method | One-step prediction (Training data) | | Long-term prediction (Training data) | | Long-term prediction (Test data) | | AIC | BIC | $N$ |
|---|---|---|---|---|---|---|---|---|---|
| | $y_1$ | $y_2$ | $y_1$ | $y_2$ | $y_1$ | $y_2$ | | | |
| Original SINDy | 0.994 | 0.995 | 0.986 | 0.994 | 0.982 | 0.995 | −2613 | −2587 | 8 |
| Dropout-SINDy | 0.995 | 0.994 | 0.987 | 0.994 | 0.984 | 0.994 | −2620 | −2570 | 15 |

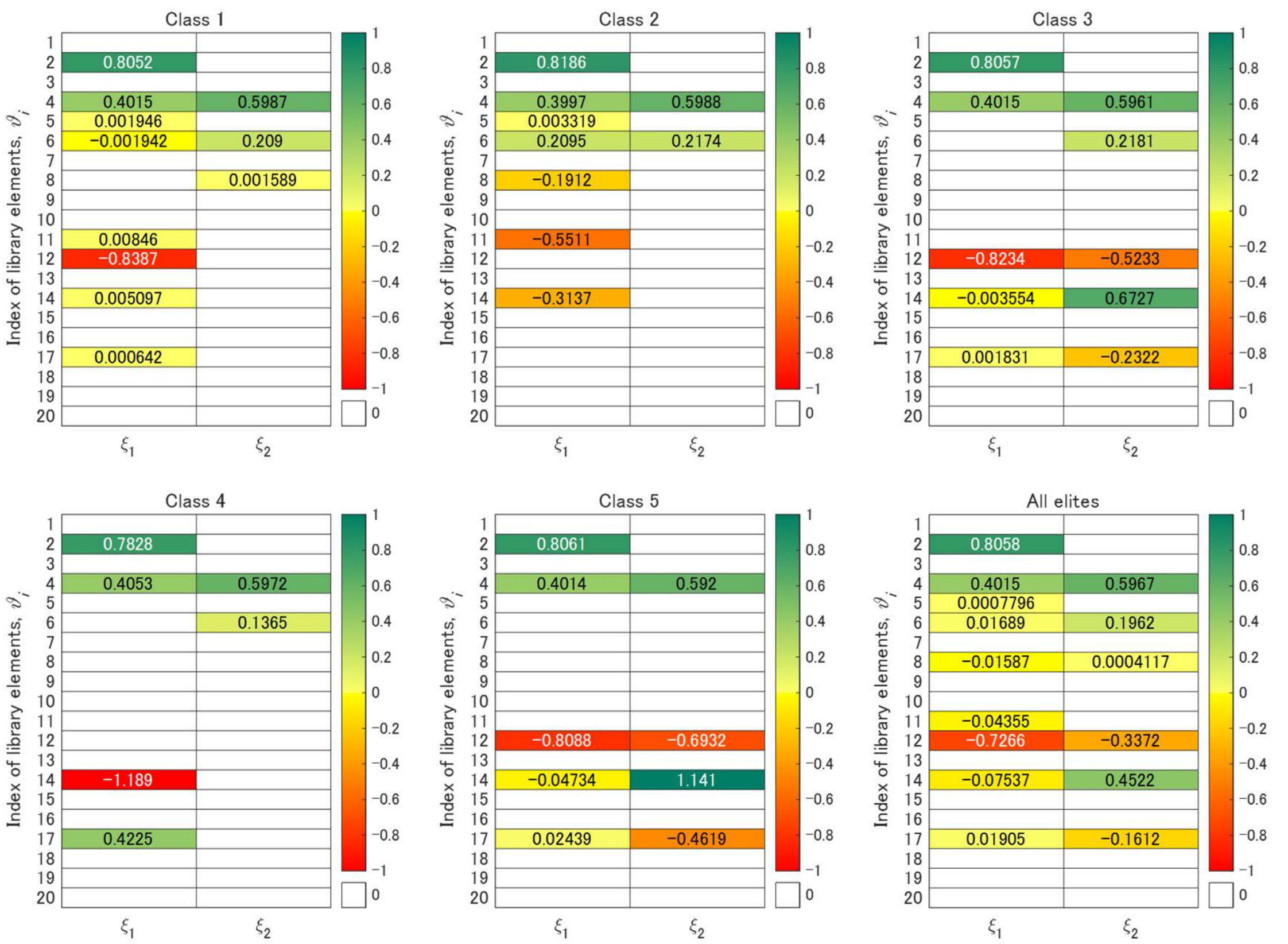

Fig. 5. Identified coefficient matrix $\Xi$ for simple example under noiseless case. Class 2 is the best model in terms of R-squares, AIC, and BIC.

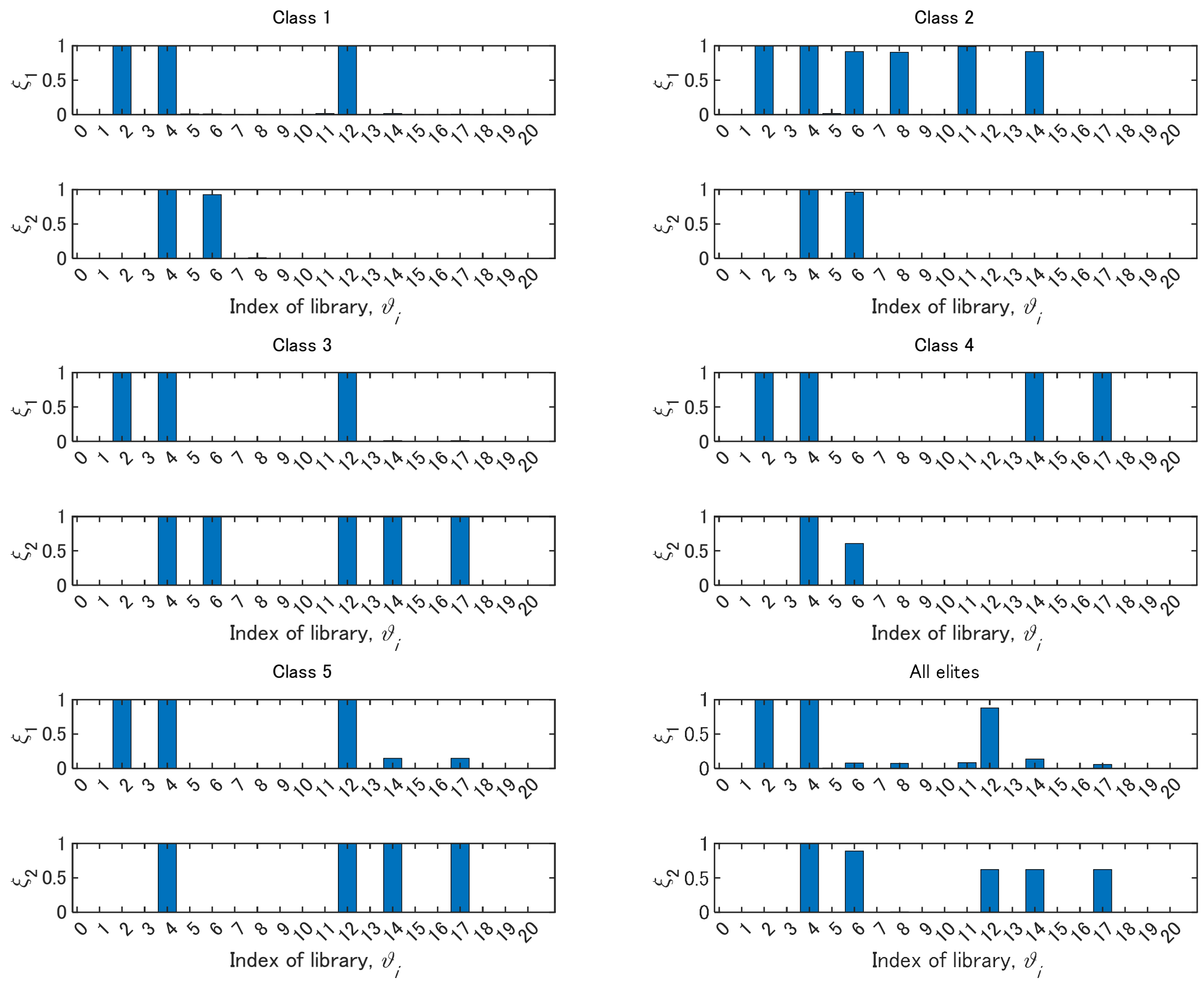

Fig. 6. Inclusion probability for simple example under noiseless case. Class 2 is the best model in terms of R-squares, AIC, and BIC.

Next, we consider a noisy case. Fig. 3 shows the overview of the time-series data obtained by exciting the input under conditions (with the noise of 30 %). In the proposed method, the classification is performed for the surviving elites. *k*-means clustering is performed and divides into five clusters. We confirmed that proper clustering is realized. Various performance metrics are examined as well as the noiseless case. Table IV indicates the R-squared of one-step and multi-step predictions for training data, multi-step predictions for test data, AIC, BIC, and the number of elements of the identified coefficient matrix in the proposed method. Table V shows those metrics of the comparison methods, such as original SINDy and dropout-SINDy. From these performance metrics of R-squared, AIC, and BIC, it can be seen that the proposed method (Class *5*) provides a good model.

We discuss the proposed method, comparing the comparison methods. The elite gathering using multi-step prediction and library bugging contributes to the improvement of prediction from the results of the traditional methods (original SINDy and dropout-SINDy) and the proposed method. The clustering of the identified coefficient matrix of elites contributes to sparsity. While the results of the proposed method without clustering (denoted as "All elites" in the table) and dropout-SINDy, which do not include the clustering procedure, increase the number of elements of the identified coefficient matrix, the proposed method with a clustering procedure provides more sparse identification.

Based on the above, the proposed method is considered to have achieved the best performance across all evaluation metrics.

TABLE IV. THE RESULTS OF R-SQUARED AND THE NUMBER OF PARAMETERS $N$ IN THE PROPOSED METHOD.

| The proposed method | One-step prediction (Training data) | | Long-term prediction (Training data) | | Long-term prediction (Test data) | | AIC | BIC | $N$ |
|---|---|---|---|---|---|---|---|---|---|
| | $y_1$ | $y_2$ | $y_1$ | $y_2$ | $y_1$ | $y_2$ | | | |
| Class 1 | 0.971 | 0.983 | 0.977 | 0.982 | 0.975 | 0.987 | −2385 | −2359 | 8 |
| Class 2 | 0.966 | 0.979 | 0.962 | 0.978 | 0.970 | 0.985 | −2323 | −2300 | 7 |
| Class 3 | 0.971 | 0.982 | 0.977 | 0.982 | 0.975 | 0.987 | −2388 | −2372 | 5 |
| Class 4 | 0.966 | 0.981 | 0.962 | 0.980 | 0.970 | 0.986 | −2340 | −2324 | 5 |
| Class 5 | 0.966 | 0.993 | 0.962 | 0.992 | 0.970 | 0.991 | −2418 | −2392 | 8 |
| All elites | 0.971 | 0.985 | 0.974 | 0.983 | 0.975 | 0.988 | −2389 | −2352 | 11 |

TABLE V. THE RESULTS OF R-SQUARED, AIC, BIC AND THE NUMBER OF COEFFICIENT PARAMETERS $N$ IN THE STANDARD METHOD.

| The comparison method | One-step prediction (Training data) | | Long-term prediction (Training data) | | Long-term prediction (Test data) | | AIC | BIC | $N$ |
|---|---|---|---|---|---|---|---|---|---|
| | $y_1$ | $y_2$ | $y_1$ | $y_2$ | $y_1$ | $y_2$ | | | |
| Original SINDy | 0.971 | 0.979 | 0.976 | 0.978 | 0.975 | 0.985 | −2359 | −2336 | 7 |
| Dropout-SINDy | 0.971 | 0.985 | 0.974 | 0.983 | 0.975 | 0.988 | −2389 | −2352 | 11 |

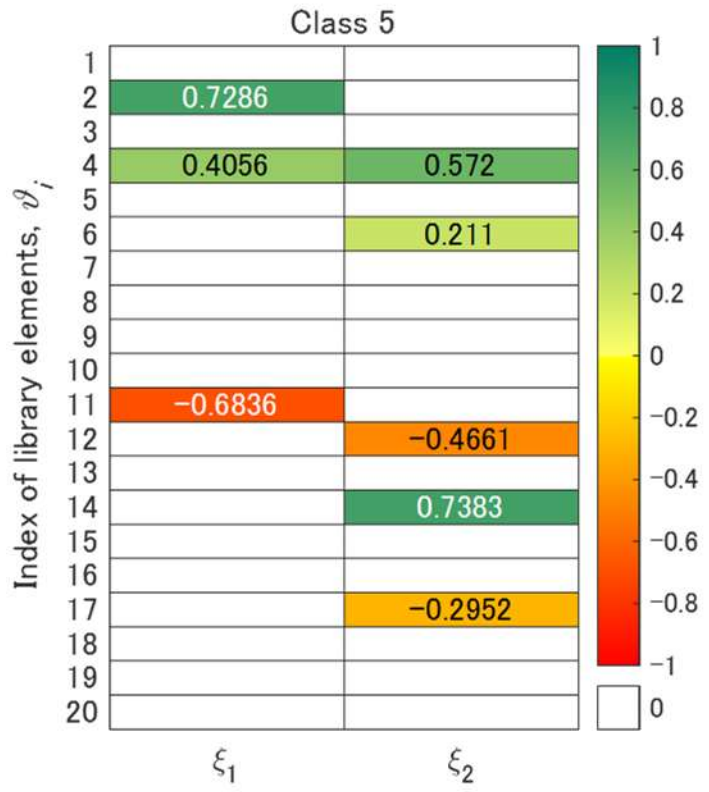

Fig. 7. Identified coefficient matrix $\Xi$ of Class 5 for simple example under noisy case.

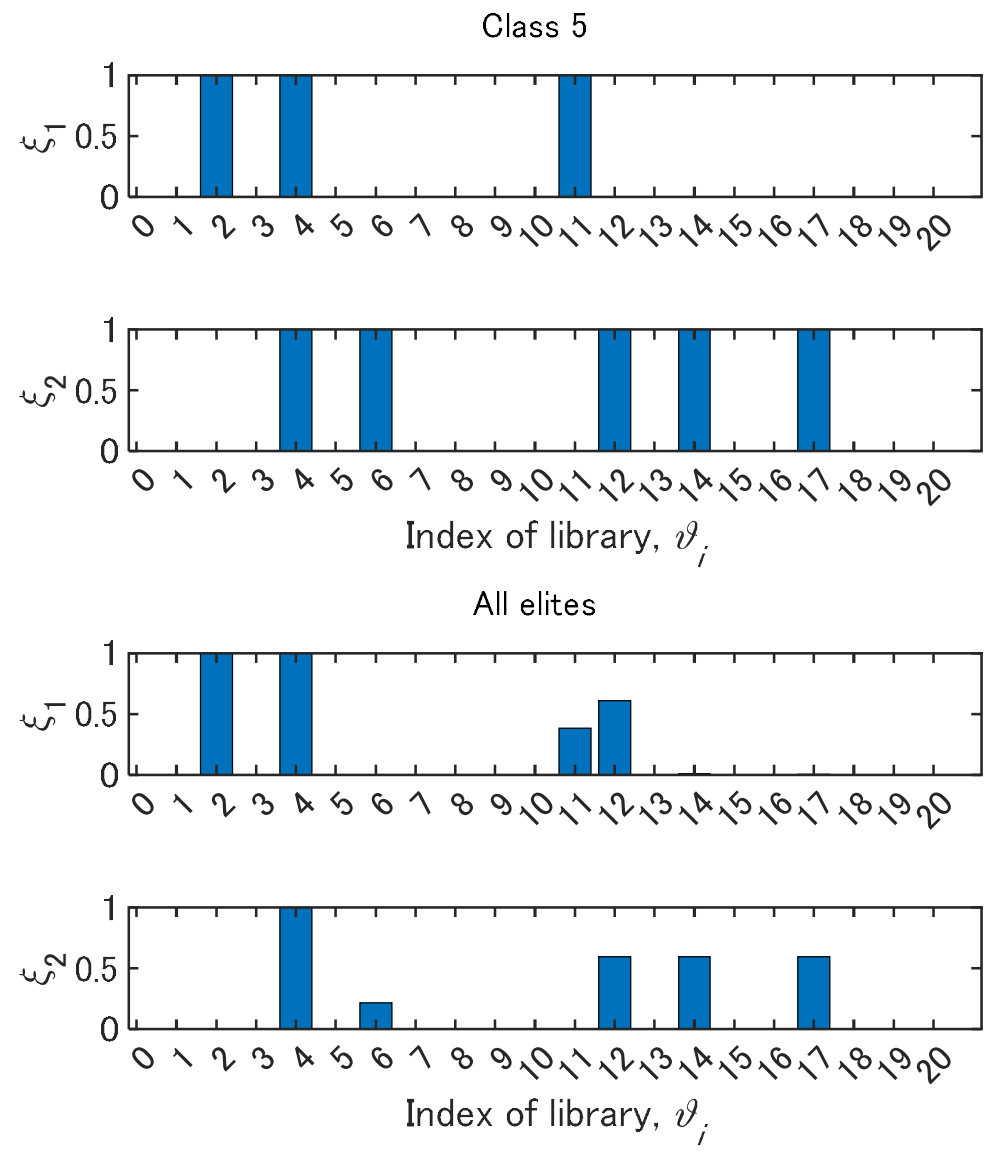

Fig. 8. Inclusion probability for simple example (Class 5 and all elites) under noisy case. is the best model in terms of R-squared, AIC, and BIC.

## 4.2.Application to diesel engine airpath system

### 4.2.1 Simulation setting

This section presents a simulation-based application of the proposed EMEC-SINDy to a diesel engine airpath system. The simulation model of the airpath system uses the mean value engine model (Wahlström and Eriksson, 2011). In this paper, we used the same simulation model as the one presented in the literature (Yahagi et al., 2025). The user-defined time-delay order of the state is set to $\sigma_x = 1$. That is, time-delay embedding is defined as $x_m(t) = [x(t) \quad x(t-1)]^T$ and the states $x$ adopt outputs $y$. The basic and proposed SINDy employ $\lambda = 1$, which is the sparse promoting parameter. The sampling period is set to 0.1s based on a prior study (Moriyasu et al., 2019). The library used prior to bagging is defined as a quadratic polynomial:

$$\vartheta = [\vartheta_1 \quad \vartheta_2 \quad \cdots \quad \vartheta_i \quad \cdots \quad \vartheta_{45}] \tag{26}$$

where $\vartheta_1 = 1, \vartheta_2 = y_1, \vartheta_3 = y_2, \vartheta_4 = y_{1(-)}, \vartheta_5 = y_{2(-)}, \vartheta_6 = d_1, \vartheta_7 = d_2, \vartheta_8 = u_1, \vartheta_9 = u_2, \vartheta_{10} = y_1^2, \vartheta_{11} = y_1 y_2, \vartheta_{12} = y_1 y_{1(-)}, \vartheta_{13} = y_1 y_{2(-)}, \vartheta_{14} = y_1 d_1, \vartheta_{15} = y_1 d_2, \vartheta_{16} = y_1 u_1, \vartheta_{17} = y_1 u_2, \vartheta_{18} = y_2^2, \vartheta_{19} = y_2 y_{1(-)}, \vartheta_{20} = y_2 y_{2(-)}, \vartheta_{21} = y_2 d_1, \vartheta_{22} = y_2 d_2, \vartheta_{23} = y_2 u_1, \vartheta_{24} = y_2 u_2, \vartheta_{25} = y_{1(-)}^2, \vartheta_{26} = y_{1(-)} y_{2(-)}, \vartheta_{27} = y_{1(-)} d_1, \vartheta_{28} = y_{1(-)} d_2, \vartheta_{29} = y_{1(-)} u_1, \vartheta_{30} = y_{1(-)} u_2, \vartheta_{31} = y_{2(-)}^2, \vartheta_{32} = y_{2(-)} d_1, \vartheta_{33} = y_{2(-)} d_2, \vartheta_{34} = y_{2(-)} u_1, \vartheta_{35} = y_{2(-)} u_2, \vartheta_{36} = d_1^2, \vartheta_{37} = d_1 d_2, \vartheta_{38} = d_1 u_1, \vartheta_{39} = d_1 u_2, \vartheta_{40} = d_1 d_2, \vartheta_{41} = d_2 u_1, \vartheta_{42} = d_2 u_2, \vartheta_{43} = u_1^2, \vartheta_{44} = u_1 u_2, \vartheta_{45} = u_2^2$ . Here, $y_{(-)}$ denotes $y(t-1)$.

### 4.2.2 Results and discussions

Fig. 9 shows the overview of the time-series data obtained by exciting control and exogenous inputs under noiseless conditions. The data from 0 to 2500 seconds is used for training, and the data from 2500 to 5000 seconds is used for testing. Fig. 10 shows the enlarged view between 1250 s and 1260s of the time-series data, where the horizontal axis denotes time and the vertical axis denotes $y_1$: boost pressure [kPa], $y_2$: EGR ratio [%], $u_1$: VGT vane closure [%], $u_2$: EGR valve opening position [%], $d_1$: fuel injection quantity [mm$^3$/st], $d_2$: engine revolution [rpm]. As preprocessing, centering is performed. Using this data, the original SINDy is first considered to obtain the ODE model. Then, the R-squared of one-step predictions of $y_1$ and $y_2$ are 0.988 and 0.991, respectively. The prediction results for training data are shown in Fig. 11. This figure depicts the simulation results of the long-term prediction with initial states and given inputs when using the original SINDy and dropout-SINDy. This figure shows that the response becomes unstable around 200 s. From this fact, we can see that outputs become unstable even if the R-squared of one-step predictions is a high score.

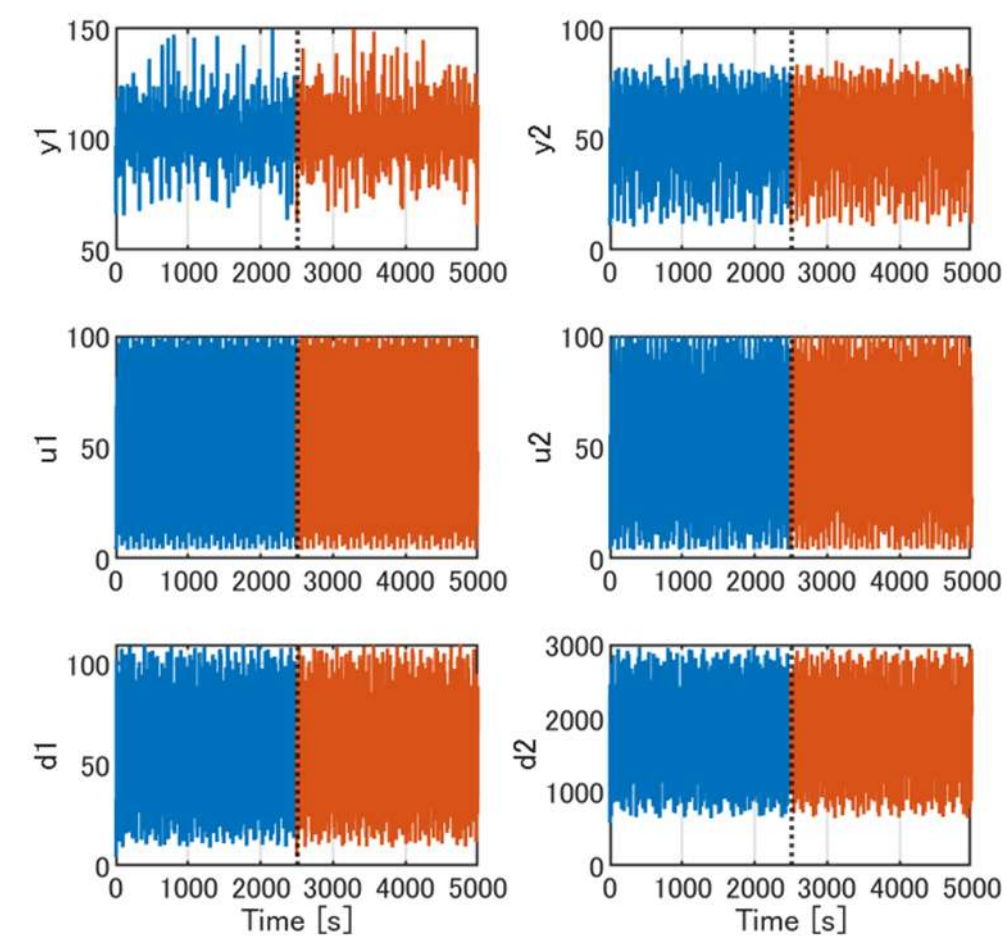

Fig. 9. Measurement data obtained by exciting control and exogenous inputs under noiseless conditions. $y_1$: boost pressure [kPa], $y_2$: EGR ratio [%], $u_1$: VGT vane closure [%], $u_2$: EGR valve opening [%], $d_1$: fuel injection amount [mm$^3$/st], $d_2$: engine revolution [rpm].

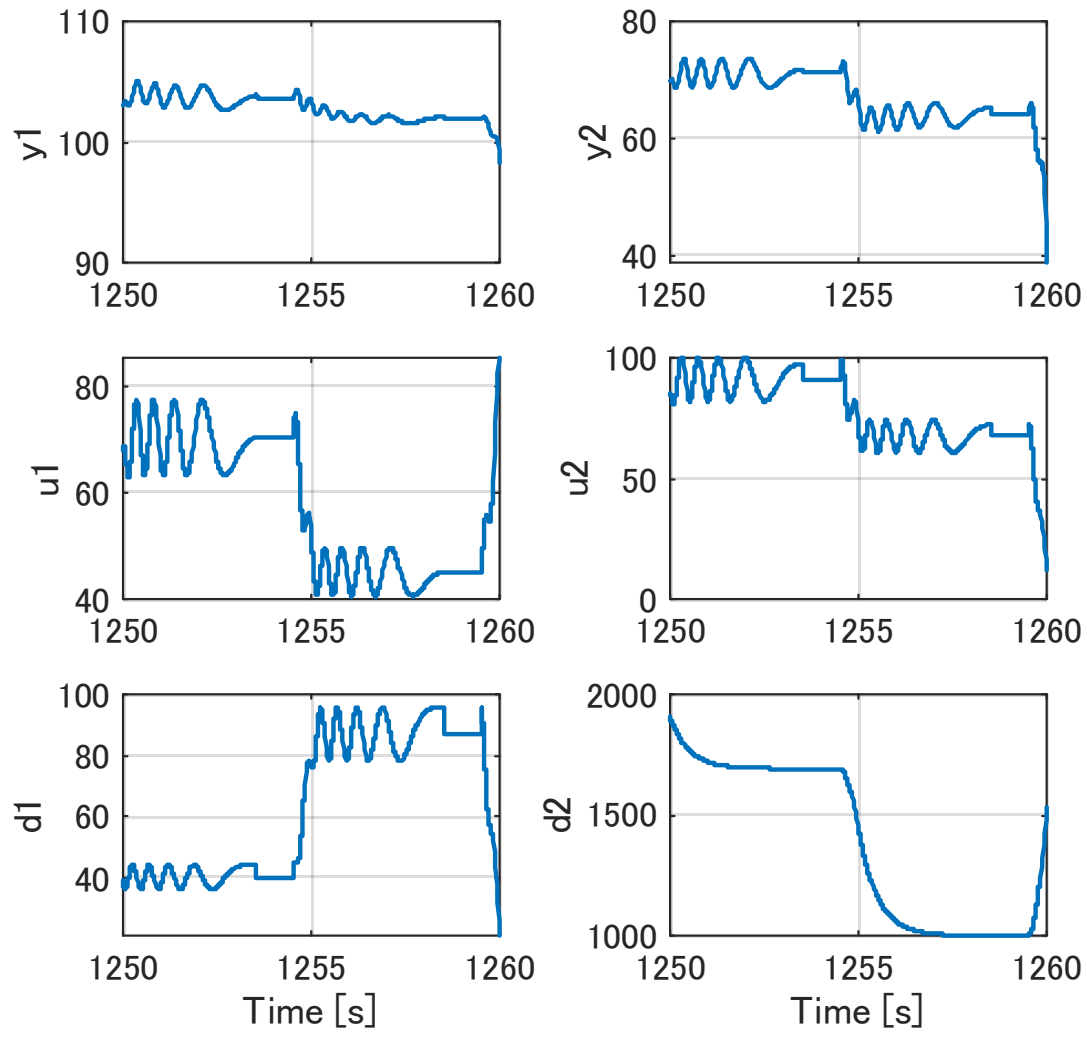

Fig. 10. Enlarged view of measurement data shown in Fig. 9. $y_1$: boost pressure [kPa], $y_2$: EGR ratio [%], $u_1$: VGT vane closure [%], $u_2$: EGR valve opening position [%], $d_1$: fuel injection quantity [mm$^3$/st], $d_2$: engine revolution [rpm].

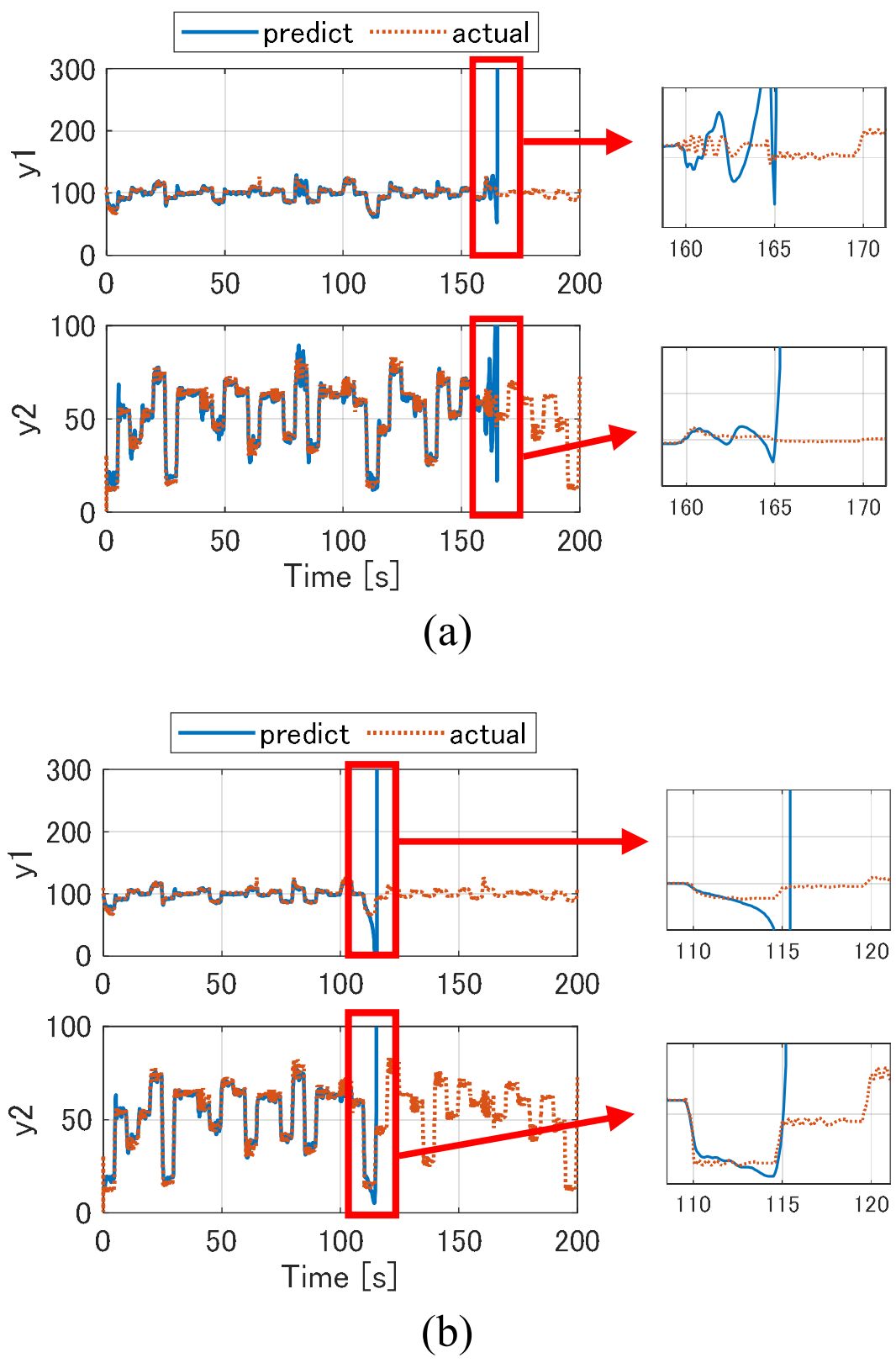

Fig. 11. Simulation results of conventional methods for training data (a) Original SINDy; (b) Dropout-SINDy. $y_1$: boost pressure [kPa], $y_2$: EGR ratio [%].

Next, the effectiveness of the proposed method is verified under ideal conditions. Based on Algorithm 1, the elites whose R-squared is more than 90 % are first collected. The calculation time and the number of iterations to gather the 50 elites were 164 s and 415, respectively. In dropout-SINDy (Abdullah et al., 2022), the number of models used in ensemble learning is set to 10. In this study, the number of elites was set to 50, considering clustering. The classification is performed for the surviving elites. Clustering was performed under the same conditions as in Section 4.1.2. Fig. 12 shows the classification results. The vertical axis represents the cluster, and the horizontal axis represents the silhouette value. We can see that proper clustering is realized because all silhouette values are positive. Fig. 13 depicts the time-series response with the proposed SINDy. The figure (a) and (b) show an overview of the data and the part around the overall maximum errors. The horizontal axis represents time, and the vertical axis represents outputs and prediction errors of $y_1$ and $y_2$. Although there are some errors, we can see that a good fit has been achieved. From the figure, the proposed method achieved stable simulation, which was not possible with conventional methods. In the elite extraction step, we confirmed that using elites selected based on one-step prediction performance, rather than multi-step prediction, resulted in models that failed to achieve accurate long-term predictions. Additionally, the effectiveness of the proposed method is evaluated using various criteria. Table VI shows the performance metrics, including R-squared, AIC, and BIC, based on the mean values for each class and for all elites. *N* denotes the number of non-zero coefficients, indicating sparsity for training and test data. To demonstrate the effectiveness of this method, we compare traditional methods such as original SINDy, dropout-SINDy, and NARX. Table VII shows the results in the comparison methods. The configuration of NARX models (NARX 1 and NARX 2) is defined as follows: NARXs use an output order of 2 and an input order of 2 with an input delay of 1, resulting in regressors such as $y(t-1), y(t-2), u(t-1), u(t-2), d(t-1), d(t-2)$. The neural network architecture of NARX 1 and NARX 2 consists of a single hidden layer with 5 and 10 neurons, respectively. The number of neurons in the hidden layer was set to correspond to the order of the model derived using SINDy, in order to ensure a fair comparison of model complexity. In the training, the Levenberg-Marquardt algorithm, which does not use learning rate setting, is used for 1000 epochs. Model selection was based on multiple criteria, including mean squared error (MSE). The NARX models were designed using the MATLAB function "narxnet". To maintain consistency with the SINDy-derived model, we initially adopted the same set of regressors in the NARX model. However, due to a substantial drop in prediction accuracy, we extended the regressor set to include $u(t-2), d(t-2)$. Original SINDy and dropout-SINDy provide unstable responses for long-term prediction, whereas the surviving all elites provide highly accurate performance for long-term prediction. The R-squared of NARX 1 models was less than 0.9 and the predictive output of NARX 1 models diverged. Additionally, the proposed method provides higher performance than others in terms of R-squared, AIC and BIC. Thus, the proposed elite's extraction using multi-step prediction is effective. Based on the above metrics, Class 2 model generated by the proposed method is selected. Therefore, the proposed method provides the best performance across all evaluation metrics.

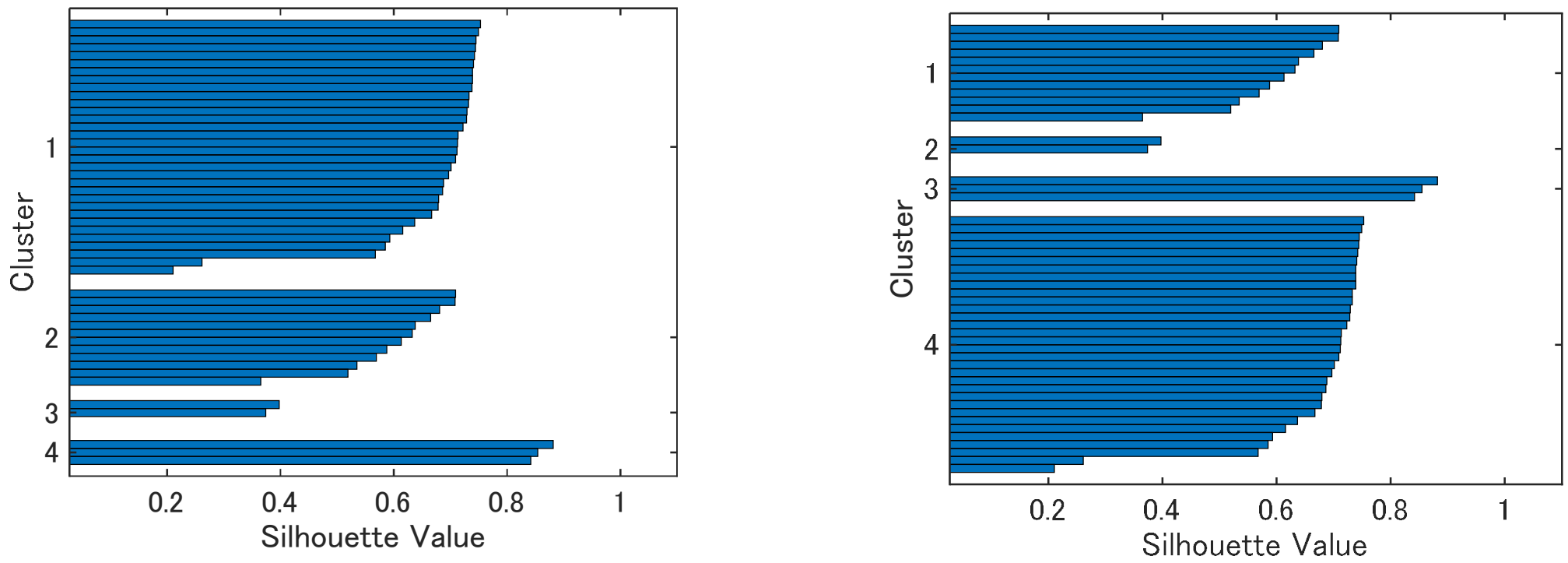

(a) *k*-means (b) Hierarchical clustering

Fig. 12. The results of *k*-means and hierarchical clustering for diesel engine airpath system.

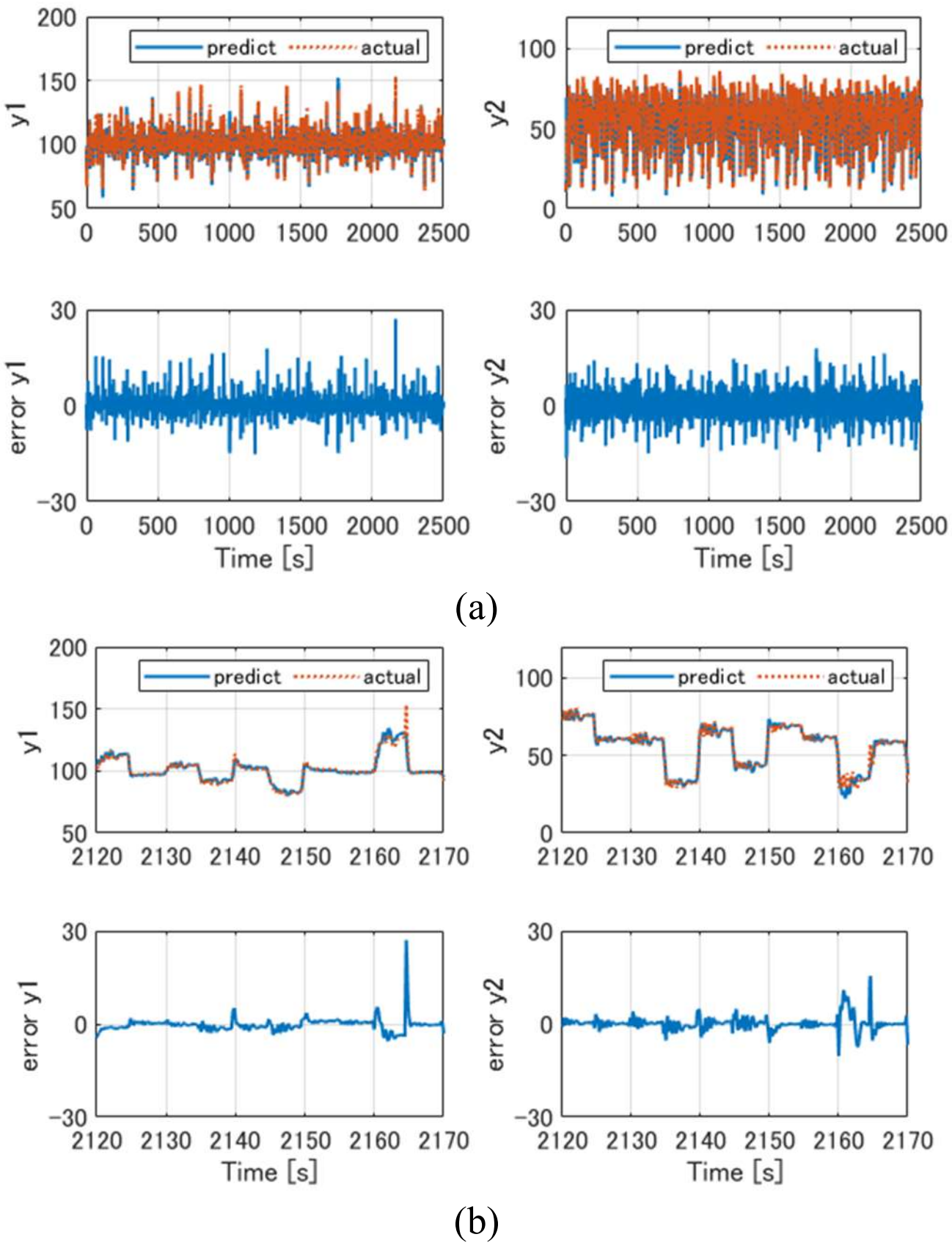

Fig. 13. Simulation results of time-series data predicted by the proposed SINDy model: (a) Overview and (b) Enlarged view of the maximum error part. $y_1$: boost pressure [kPa], $y_2$: EGR ratio [%].

TABLE VI. THE RESULTS OF R-SQUARED, AIC, BIC AND THE NUMBER OF COEFFICIENT PARAMETERS $N$ IN THE PROPOSED METHOD.

| The proposed method | One-step prediction (Training data) | | Long-term prediction (Training data) | | Long-term prediction (Test data) | | AIC | BIC | $N$ |
|---|---|---|---|---|---|---|---|---|---|
| | $y_1$ | $y_2$ | $y_1$ | $y_2$ | $y_1$ | $y_2$ | | | |
| Class 1 | 0.989 | 0.992 | 0.937 | 0.967 | 0.939 | 0.966 | 67.9 | 377 | 85 |
| Class 2 | 0.989 | 0.992 | 0.959 | 0.982 | 0.961 | 0.982 | −278 | −1.28 | 76 |
| Class 3 | 0.989 | 0.993 | 0.953 | 0.966 | 0.956 | 0.966 | −19.3 | 274 | 70 |
| Class 4 | 0.990 | 0.995 | 0.928 | 0.990 | 0.928 | 0.990 | 545 | 799 | 70 |
| All elites | 0.989 | 0.992 | 0.952 | 0.978 | 0.954 | 0.978 | −140 | 172 | 86 |

TABLE VII. THE RESULTS OF R-SQUARED, AIC, BIC AND THE NUMBER OF COEFFICIENT PARAMETERS $N$ IN THE STANDARD METHOD.

| The comparison method | One-step prediction (Training data) | | Long-term prediction (Training data) | | Long-term prediction (Test data) | | AIC | BIC | $N$ |
|---|---|---|---|---|---|---|---|---|---|
| | $y_1$ | $y_2$ | $y_1$ | $y_2$ | $y_1$ | $y_2$ | | | |
| Original SINDy | 0.988 | 0.991 | 0.988 | 0.991 | Unstable (−1.17) | Unstable (−1.45) | 994000 | 995000 | 87 |
| Dropout-SINDy | 0.990 | 0.993 | 0.990 | 0.993 | Unstable (−1.16) | Unstable (−1.44) | 828000 | 829000 | 84 |
| NARX 1 | 0.989 | 0.984 | 0.571 | 0.718 | 0.603 | 0.726 | 127000 | 127000 | 57 |
| NARX 2 | 0.993 | 0.988 | −855 | −421 | −848 | −418 | 396000 | 396000 | 112 |

Fig. 14 presents the heatmap of the identified coefficient matrix $\Xi$ for each class. For clarity, the coefficient matrix $\tilde{\xi}$ is shown, which corresponds to the normalized matrix $\tilde{\Theta}$. Specifically, the relationship of the normalized and original coefficient matrices is given by $\tilde{\xi}_j = \sigma_j \xi_j$ where $\sigma_j$ denotes the norm of $\Theta_j$, $j$-th row of original matrix $\Theta$. The heatmap implies that features with significant impact in the library are effectively extracted, in contrast to neural-network-based approaches, which yield fully black-box models. Fig. 15 shows the inclusion probability (Fasel et al., 2022) for each class as well as for all elites, where $\xi_3$ and $\xi_4$ are omitted due to their results being self-evident. From the figures and Table VI, we can see that clustering contributes to sparsity, as well as the previous section. In simple examples, such as those introduced in the previous section, where the system dynamics can be expressed as a linear combination of candidate basis functions in the library, it is possible to verify whether the extracted features correspond to those of the actual system. In more complex and highly coupled nonlinear MIMO systems—such as the air path system of a diesel engine—where the appropriate basis functions are not readily known, it is difficult to conduct this validation. Note that, unlike many prior studies, this section does not assume that the system can be explicitly expressed as a first-principles model. In such cases, where physically motivated basis functions are not explicitly known, the proposed EMEC-SINDy method still provides an accurate model.

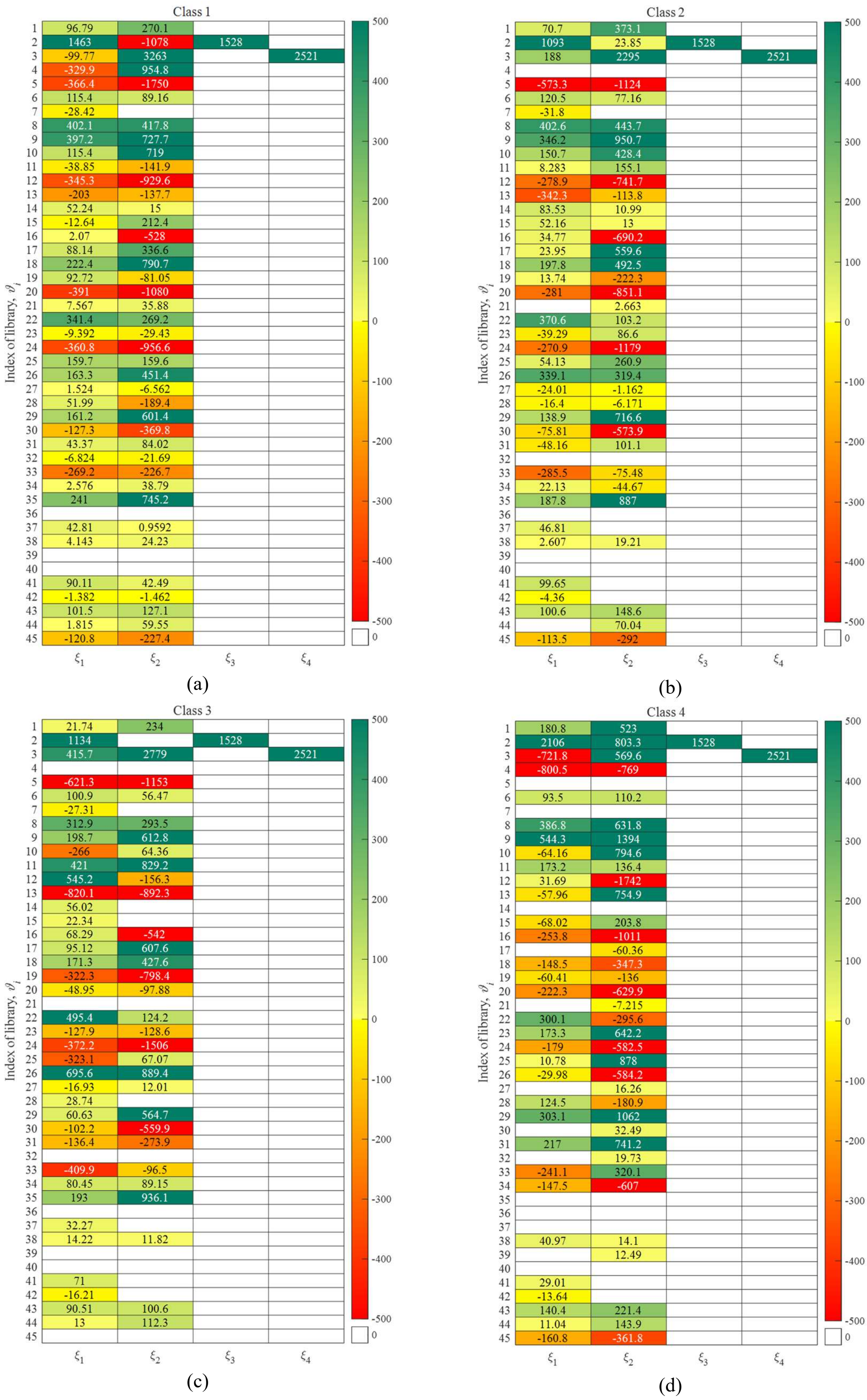

Fig. 14. Identified coefficient matrix $\Xi$ for normalized $\Theta$ ($\lambda = 30$): (a) Class 1, (a) Class 2, (a) Class 3, (a) Class 4.

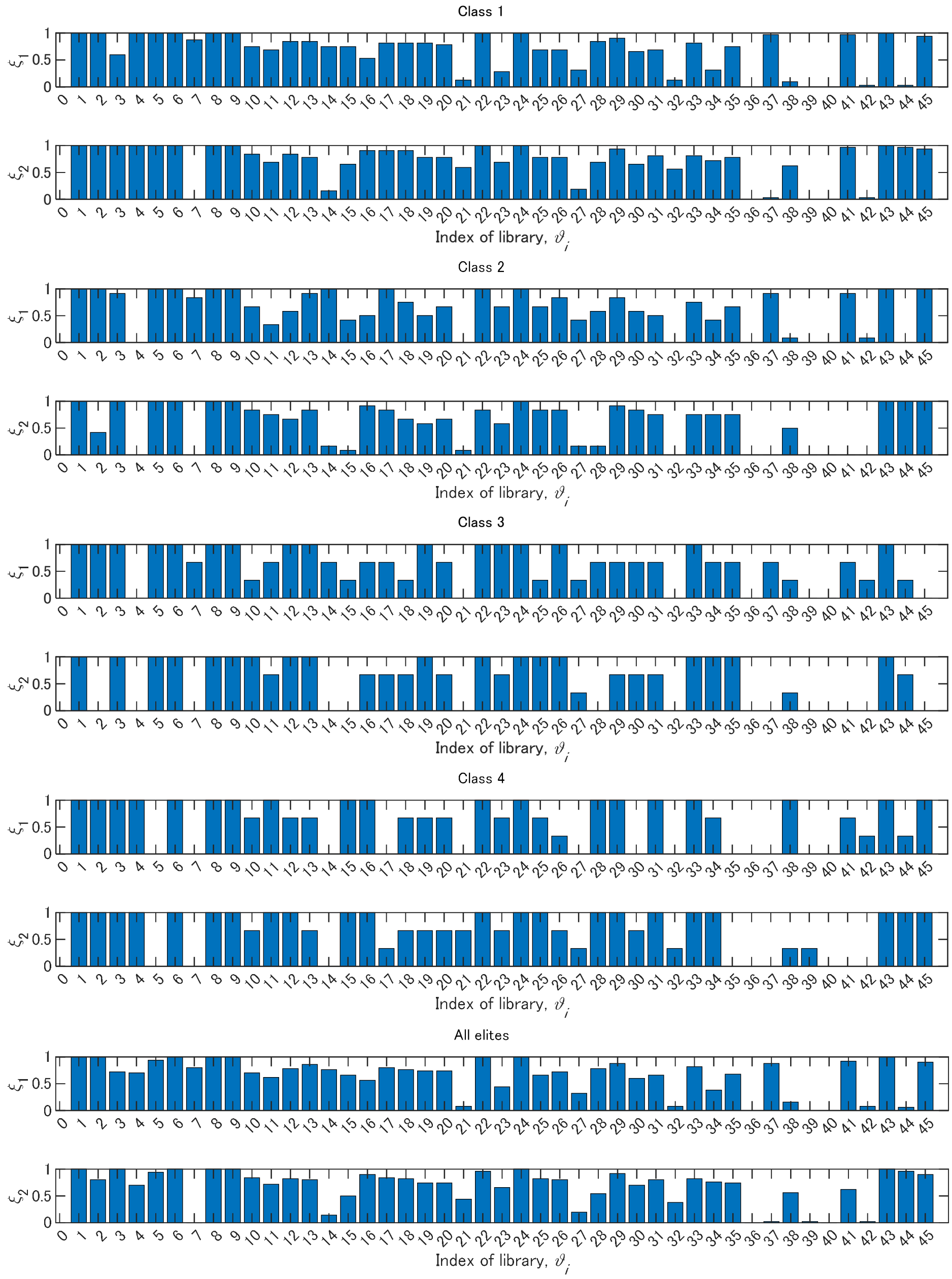

Fig. 15. Inclusion probability.

The proposed method is verified under noisy conditions considering real-world applications. We consider the noise level from 5 to 20 %. Fig. 16 shows the enlarged view of the outputs with the noise of 20 %. This figure was illustrated to visually illustrate the noise level. It is highlighted that the noise may lead to a reduction in the modeling accuracy (Fasel et al., 2022; França et al., 2022). Thus, this consideration under noisy conditions is essential. Table VIII shows the simulation results with different noise levels. In the table, the R-squared of one-step and long-term predictions, and the number of coefficients are shown. The computation time and iterations to gather the elites are also indicated. There is a variation in the computation time and the number of iterations because the library bagging is performed randomly. From the table, the proposed ensemble-based SINDy provides the discrete model to realize multi-step predictions for industrial systems under noisy conditions.

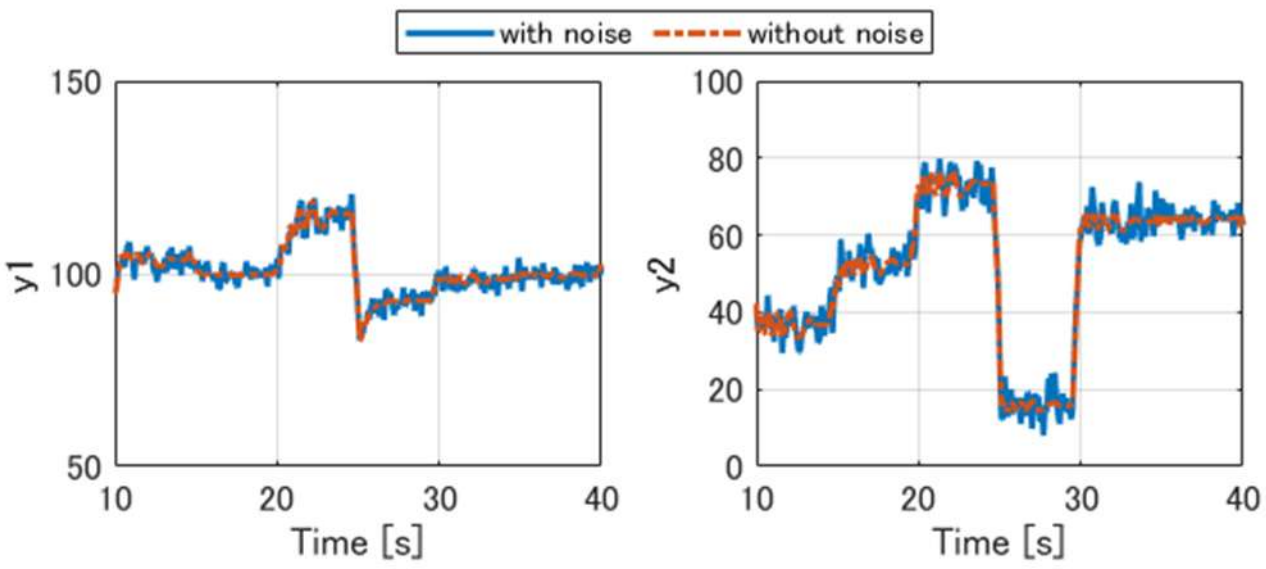

Fig. 16. Enlarged view of the outputs with 20 % noise level. $y_1$: boost pressure [kPa], $y_2$: EGR ratio [%].

Table VIII. The results with different noise levels.

| Noise [%] | R-squared of one-step prediction | | R-squared of long-term prediction | | $N$ | Computation time [s] | Iteration |
|---|---|---|---|---|---|---|---|
| | $y_1$ | $y_2$ | $y_1$ | $y_2$ | | | |
| 0 | 0.989 | 0.992 | 0.959 | 0.982 | 76 | 164 | 415 |
| 5 | 0.989 | 0.993 | 0.965 | 0.990 | 79 | 74 | 133 |
| 10 | 0.986 | 0.988 | 0.962 | 0.991 | 78 | 83 | 149 |
| 15 | 0.978 | 0.976 | 0.961 | 0.991 | 83 | 170 | 323 |
| 20 | 0.962 | 0.960 | 0.953 | 0.991 | 80 | 231 | 414 |

Finally, we discuss the difference between this study and the prior work (Yahagi et al., 2025). In the previous study, MPC combined with the original SINDy has been introduced and evaluated for the airpath system, which is the same system used in this study. The results confirmed that MPC with SINDy works effectively for the intake and exhaust system. However, the improvement of the R-squared value for long-term prediction was limited. To enhance the expressive capability of SINDy, we enriched the library by introducing time-delay coordinates. Nevertheless, due to overfitting, an appropriate model could not be obtained. As illustrated in Fig. 11, the original SINDy with time-delay coordinates resulted in significantly poor multi-step prediction performance. In contrast, this study applied EMEC-SINDy, introduced in Section 3, to the same airpath system. EMEC-SINDy, which incorporates ensemble learning, multi-step prediction evaluation, elite strategy, and classification techniques, differs fundamentally from the original SINDy. Additionally, the AIC and BIC have been evaluated in this study. The proposed method demonstrated superior fitting accuracy compared to existing approaches.

# 5. Conclusion

This paper has presented the EMEC-SINDy with inputs and the extended coordinate to achieve highly accurate and reliable predictions by utilizing ensemble learning, collection of the elites, classification techniques, and evaluation of multi-step prediction for an industrial system under noisy conditions. In the proposed method, library bagging, which is a technique of ensemble learning, is performed, leaving elites with an R-squared greater than 90% for multi-step predictions. This elite collection contributes to improving prediction accuracy for complex and highly coupled nonlinear MIMO systems such as industrial systems. Clustering is performed on the surviving elites because physically motivated basis functions are not always available, and the elite models are not necessarily expressed as a linear combination of the same basis functions. This clustering process is effective in promoting sparsity. After the classification is applied to the elites, a model is obtained by taking the mean of the final classified elites. Finally, the best model is selected. The proposed method features the realization of multi-step predictions under noisy situations without using nonlinear optimization problems. In the simulation, the proposed method is applied to the diesel engine airpath system with nonlinear and MIMO characteristics. The results show that the model identified by the proposed algorithm realizes the multi-step predictions. Thus, the proposed method is effective for the use of simulation plants and their application to model-based controllers such as MPC. Future work will focus on designing real-time implementation strategies and integrating them with embedded controllers by introducing MPC with the SINDy model derived from the proposed methodology. In addition, real-world applications to diesel engine airpath systems will be explored. In this context, we will validate both the accuracy of the SINDy model identified by the proposed algorithm and the control performance of MPC with the SINDy model.

# Appendix

This section describes the integration of SINDy model into the control problem. MPC with SINDy in the discrete-time form is formulated as the following optimization problem:

$$\min_{u(0),u(1),\cdots,u(H_u)} \sum_{i=0}^{H_p-1} \|x^r(k) - x(k+i)\|_Q^2 + \sum_{i=0}^{H_u-1} \|u(k+i)\|_{R_u}^2 + \|\Delta u(k+i)\|_{R_{\Delta u}}^2 \quad (27)$$

subject to

$$x(i+1) = \varXi\vartheta^T(x(i), u(i), d(i))$$

$$x_{min} \leq x(i) \leq x_{max}$$

$$u_{min} \leq u(i) \leq u_{max}$$

$$\Delta u_{min} \leq \Delta u(i) \leq \Delta u_{max}$$

where $k$ is the current time; $i$ is the time of prediction interval; $x^r$ is target value; $\Delta u(i) = u(i) - u(i-1)$; $Q$, $R_u$, and $R_{\Delta u}$ are positive definite diagonal matrices, and are the weights for the deviation, the integral of the deviation, the control input, and the amount of change in the control input, respectively; $\|\bullet\|_Q^2$ represents the weighted 2-norm of $Q$, defined as $\|x\|_Q^2 = x^T Q x$; $H_p$ and $H_u$ are the prediction horizon and control horizon, respectively. The Sequential Quadratic Programming (SQP) is known as a general solver for the above optimization problem. In MPC, the first control input $u(0)$ is applied to the targeted system.